\documentclass[10pt, a4paper]{article}
\usepackage[T1]{fontenc}
\usepackage{amsmath,amsfonts, amssymb, amsbsy, mathrsfs, mathtools, bbm, enumitem, cases}
\usepackage[medium]{titlesec}
\usepackage{float}
\usepackage{makeidx}
\usepackage{graphicx}
\usepackage[x11names]{xcolor}
\usepackage{url}
\usepackage{authblk}
\usepackage{multirow}
\usepackage[font=small,labelfont=bf]{caption}
\usepackage{mathpazo}
\usepackage[toc]{appendix}
\usepackage[hidelinks,colorlinks=true,linkcolor=red,citecolor=blue,urlcolor=magenta]{hyperref}
\usepackage[left=2.75cm, right=2.75cm, top=3.0cm, bottom=3.00cm]{geometry}
\usepackage{rotating}

\usepackage{fancyhdr}

\newcommand{\email}[1]{\href{mailto:#1}{#1}}

\newcommand{\df}{\textrm{d}}

\newcommand{\dprime}{{\prime\prime}}

\newcommand{\e}{\textrm{e}}
\newcommand{\hf}{{\frac{1}{2}}}

\renewcommand{\tilde}{\widetilde}

\titleformat{\section}
{\normalfont\large\bfseries}{\thesection}{1em}{}

\titleformat{\subsection}
{\normalfont\normalsize\bfseries}{\thesubsection}{1em}{}

\titleformat{\paragraph}
{\normalfont\normalsize\bfseries}{\theparagraph}{1em}{}

\titleformat{\subparagraph}
{\normalfont\normalsize\bfseries}{\thesubparagraph}{1em}{}

\newcommand\scalemath[2]{\scalebox{#1}{\mbox{\ensuremath{\displaystyle #2}}}}

\usepackage[square,numbers,sort&compress]{natbib}

\numberwithin{equation}{section}

\begin{document}
	\setlength{\bibsep}{0pt}
	
	\title{ \Large\textbf{The Tolman VII Space-time in the Presence of Charge and a Cosmological Constant}}
	\author[]{\normalsize James Ripple\thanks{\email{jfripple@smcm.edu} } }
	\author[]{\normalsize Anish Agashe\thanks{\email{anagashe@smcm.edu} (corresponding author)} }
	\affil[]{\small \it Department of Physics and Materials Science,\\ \it St. Mary's College of Maryland,\\ \it 47645 College Dr, St. Mary's City,\\ \it Maryland, USA 20686}
	
	\date{}

	\maketitle

	\begin{abstract}
		The Tolman VII space-time is one of the few physically acceptable exact solutions in general relativity. In this paper, we derive a generalised Tolman VII solution which includes a charge and a cosmological constant. We analyse the spatial geometry of the solution and present conditions for zero and non-zero spatial curvature. We show that for a particular value of the boundary, the Tolman VII space-time can be matched to the charged Nariai space-time. This represents a new class of interior Nariai solutions. Matching with the Reissner-Nordstr\"om-de Sitter space-time, we derive analytic expressions for the metric functions and the pressure. Using this, we show that the solution allows for trapped null geodesics for a broad range of values for the total charge, central density, and the cosmological constant. We investigate the physical properties and derive an equation of state for the fluid. We show that the fluid can be considered a polytrope with, $\Gamma \sim 2.5$. Finally, we analyse the sound speed and energy conditions to conclude that only a subclass of the solution follows all the basic physical acceptability criteria.\\
		
	\end{abstract}

	{
		\hypersetup{linkcolor=black}
		\tableofcontents
	}
	
	\section{Introduction}
	Static spherically symmetric fluid solutions of the Einstein field equations have been of great interest ever since the advent of general relativity (GR). These solutions commonly denote the interior space-time of a finite sized fluid sphere and hence can be used to model compact objects such as stars \cite{mtw}. The earliest such solution can be traced back to Schwarzschild who assumed a fluid with a uniform density \cite{schwint}. Many other solutions including those with a varying density have been found since. The Tolman VII solution is one such example \cite{tolman1}. It assumes a density that is quadratic in the radial coordinate. It is one of the very few physically acceptable spherically symmetric fluid solutions \cite{delg} and has been shown to be a viable candidate to model neutron stars and self-bound stars (such as stars made of quark matter) \cite{raghoo,raghoo3,jia,pos1,pos2}. Therefore, it is of interest to analyse the properties of this solution and its generalisations.
	
	In this paper, we derive a generalised form of the Tolman VII solution that includes a charge and a cosmological constant. Charged fluid spheres have been amply discussed in the literature \cite{bek,mehra1,flor} and so have been solutions with a cosmological constant \cite{weyl,stuch1,bohm}. However, solutions with both charge and a cosmological constant have been given relatively less attention. In \cite{bohm2}, uniform density charged fluids in the presence of a cosmological constant were considered. The aim of this paper is to present a similar but much more extended analysis for the Tolman VII space-time\footnote{Tolman, in his paper \cite{tolman1}, did initially include a cosmological constant but ignored it mentioning that it is ``\textit{too small to produce appreciable effects}''. Also, see \cite{raghoo2} for Tolman VII solution with a charge.}.
	
	We start our analysis with describing the Einstein-Maxwell field equations with a cosmological constant. Then, we choose an `appropriate' form for the charge distribution and assume the Tolman VII density profile. Using this, we solve the field equations and derive the generalised form of the Tolman VII solution. Both the geometric and physical properties of the solution now depend on a parameter space that contains five parameters: central density, boundary radius, total charge, cosmological constant, and the self-boundedness parameter. We analyse the spatial curvature and obtain conditions for which the spatial hypersurfaces are curved (positive or negative) and flat. For a particular choice of the central density and the total charge, the spatial curvature becomes identical to the case of an uncharged uniform density sphere. For physically reasonable values of the five parameters mentioned above, the fluid sphere exhibits a positively curved spatial geometry. 
	
	Then, we consider matching the interior and the exterior space-times. The presence of a charge and a cosmological constant means that the exterior space-time can be described by either the Reissner-Nordstr\"om-de Sitter (charged Kottler) metric or the charged Nariai metric. We show that the generalised Tolman VII space-time can be matched to the charged Nariai space-time if the matching is done at a specific value of the boundary radius. This has been previously shown for uniform density spheres \cite{bohm2}. Here we extend this to the Tolman VII solution, thus deriving a new class of solutions interior to charged Nariai. 
	
	By matching with the Reissner-Nordstr\"om-de Sitter metric, we solve for all the unknown constants that arise from integrating the field equations. This gives us complete expressions for the metric functions and the pressure. Then, we solve the geodesic equations and derive a generalised condition for the trapping of null geodesics. We use the Tolman-Oppenheimer-Volkoff equation to convert this into a condition on the pressure and density of the fluid. We show that for a given boundary radius, the solution allows for trapped null orbits for a range of values of the rest of the parameters. We explore the full parameter space to find that depending on the values of other parameters, trapped null geodesics exist for a tenuity in the range $1.93$ to $3.07$. This extends the bounds on tenuity found in previous studies such as \cite{ishak,stuch2}. 
	
	Further, we solve for the radial profile of the pressure and obtain an analytic expression for the equation of state. Assuming the fluid to be a polytrope, we perform numerical fitting to find the values for the polytropic index that correspond the best to the analytic expression. We do this for different values of the self-boundedness parameter. Finally, we investigate the energy conditions and the sound speed to check the physical acceptability of the solution. We show that the weak energy condition ensures a positive spatial curvature. From this analysis, we conclude that only a subclass of the generalised Tolman VII solution is physically acceptable.
	
	The paper is arranged in the following manner: in section \ref{sec-fieldeqs}, we present the mathematical setup necessary to solve for the metric functions and the pressure. We solve these equations with the Tolman VII density ansatz and present geometric properties of the solution in section \ref{sec-tol7}. In section \ref{sec-extst}, we match the interior solution to different choices of exterior space-time. Section \ref{sec-geodestrap} presents an exploration the full parameter space to show that the solution allows of trapping of null geodesics. In section \ref{sec-physprop}, we present the equation of state for the fluid and investigate the physical acceptability conditions. Finally, in section \ref{sec-sumdisc}, we summarise the results in this paper and provide some concluding remarks.
	
	We work with geometric units, i.e., $G = 1,\ c=1,\ \epsilon_0 = 1$ and follow the Landau-Lifshitz space-like convention (LLSC) \cite{mtw}.

	\section{Charged Perfect Fluid Spheres with a Cosmological Constant} \label{sec-fieldeqs}
	The line element for a static spherically symmetric space-time in spherical coordinates $(t,r,\theta,\phi)$ is given by \cite{stephani,goenner},
	\begin{equation}\label{ssmetric}
		\df s^2 = -\e^{2 \Phi(r)}\df t^2 + \e^{2\Psi(r)}\df r^2 + r^2 \left( \df \theta^2 + \sin^2\theta\df\phi^2 \right)
	\end{equation}
	The time-like unit 4-vector field admitted by such space-time is given by, $u^\alpha = \left[\e^{-\Phi},0,0,0\right] $. The energy-momentum tensor for a perfect fluid is given by,
	\begin{equation}\label{emtensor}
		^{(\rm matter)}{T^\alpha}_\beta  =  (\rho + p) u^\alpha u_\beta + p \delta^\alpha_\beta
	\end{equation}
	where, $\rho\equiv\rho(r)$ and $p\equiv p(r)$ are, respectively, the energy density and pressure of the fluid. The fluid is taken to be charged such that it generates a static radial electric field, $E(r)$. The energy-momentum tensor for the electromagnetic field is given by,
	\begin{equation}\label{ememtensor}
		^{(\rm emfield)}{T^\alpha}_\beta = \frac{1}{4\pi} \left(F^{\alpha\sigma}F_{\beta\sigma} - \frac{1}{4}\delta^\alpha_\beta F^{\epsilon\sigma}F_{\epsilon\sigma} \right)
	\end{equation}
	where, due to spherical symmetry and staticity, we have,
	\begin{equation}\label{faraten}
		F_{\epsilon\sigma} = 2 \delta^1_{[\epsilon}\delta^0_{\sigma]} E(r)
	\end{equation}
	The charge density can be found using,
	\begin{equation}
		J^0 = \frac{1}{4\pi}\nabla_\sigma F^{0\sigma} 
	\end{equation}
	Using this, the charge distribution is given by,
	\begin{equation}\label{chargedist}
		q(r) = \int \sqrt{-g}\ J^0 \df^3 x = r^2 \e^{-(\Phi + \Psi)} E(r)
	\end{equation}
	where, $g = \det(g_{\alpha\beta})$.
	
	The Einstein-Maxwell field equations with a cosmological constant are given by,
	\begin{equation}\label{efelam}
		{R^\alpha}_\beta - \hf R \delta^\alpha_\beta = 8\pi \left(^{(\rm matter)}{T^\alpha}_\beta + ^{(\rm emfield)}{T^\alpha}_\beta\right) + \Lambda \delta^\alpha_\beta
	\end{equation}
	Using equations \eqref{emtensor} - \eqref{chargedist}, the field equations for the metric in \eqref{ssmetric} take the form,
	\begin{subequations}\label{genefe}
		\begin{equation}\label{efe1}
			\frac{\e^{-2\Psi}}{r^2} \left(1 - 2r\Psi^\prime \right) - \frac{1}{r^2} = -8\pi\rho - \frac{q^2}{r^4} +  \Lambda 
		\end{equation}
		
		\begin{equation}\label{efe2}
			\frac{\e^{-2\Psi}}{r^2} \left(1 + 2r \Phi^\prime\right) - \frac{1}{r^2} = 8\pi p - \frac{q^2}{r^4} + \Lambda
		\end{equation}
		\begin{equation}\label{efe3}
			\frac{\e^{-2\Psi}}{r^2}\left[ r\left( \Phi^\prime - \Psi^\prime\right) + r^2\left( \Phi^\dprime + {\Phi^\prime}^2 - \Phi^\prime \Psi^\prime \right)\right] = 8\pi p + \frac{q^2}{r^4} + \Lambda
		\end{equation}
	\end{subequations}
	where, `dot' and `prime' represent differentiation with respect to the coordinates, $t$ and $r$, respectively. The variables, $\Phi,\Psi,\rho,p,q$ are functions of $r$, while $\Lambda$ is a constant.
	
	The conservation of energy-momentum now becomes,
	\begin{equation}
		\nabla_\alpha \left(^{(\rm matter)}{T^\alpha}_\beta + ^{(\rm emfield)}{T^\alpha}_\beta\right) = 0
	\end{equation}
	This leads to only one equation (for $\beta~=~1$), while other components are identically zero,
	\begin{equation}\label{conseq}
		p^\prime + \Phi^\prime\left(\rho + p\right) = \frac{q q^\prime}{4\pi r^4}
	\end{equation}  
	
	\subsection{Solving for the Metric Functions}
	To specify the space-time completely, one needs to find the explicit functional form of the metric coefficients. This is done by solving Einstein field equations. The first field equation \eqref{efe1} can be simplified as,
	\begin{equation}
		\left[r\left(\e^{-2\Psi} - 1\right)\right]^\prime = -8\pi\rho r^2 - \frac{q^2}{r^2} +  \Lambda r^2 
	\end{equation}
	This can be integrated to give,
	\begin{equation}\label{psieq1}
		\e^{2\Psi(r)} = \left[1 - \frac{2m(r)}{r} - \frac{\epsilon(r)}{r} + \frac{\Lambda r^2}{3}\right]^{-1}
	\end{equation}
	Here, we have defined,
	\begin{subequations}
		\begin{equation}\label{massfunc}
			m(r) := 4\pi \int \rho(r) r^2\ \df r
		\end{equation}
		\begin{equation}
			\epsilon(r) := \int \frac{q^2(r)}{r^2}\ \df r
		\end{equation}
	\end{subequations}
	This remains true for any given density profile or charge distribution. Therefore, a general spherically symmetric charged fluid space-time has the following line element,
	\begin{equation}\label{ssle}
		\df s^2 = -\e^{2 \Phi(r)}\df t^2 + \frac{\df r^2}{1 - \frac{2m(r)}{r} - \frac{\epsilon(r)}{r} + \frac{\Lambda r^2}{3}} + r^2 \left( \df \theta^2 + \sin^2\theta\df\phi^2 \right)
	\end{equation}
	
	Solving for the other metric coefficient is much more complicated than this. To do this, we first use the isotropy of the pressure to get, ${G^1}_1 - {G^2}_2 + \frac{2q^2}{r^4} = 0$. This gives the following equation,
	\begin{equation}
		\e^{-2\Psi}\left[\left(1+r\Phi^\prime\right)\left(1+r\Psi^\prime\right) - r^2\left(\Phi^\dprime + {\Phi^\prime}^2\right)\right] - 1 + \frac{2q^2}{r^2} = 0
	\end{equation}
	Using this, we get,
	\begin{equation}\label{phieq1}
		\e^{-2\Psi}r^2\left(\Phi^\dprime + {\Phi^\prime}^2\right) - \e^{-2\Psi}\left(1+r\Psi^\prime\right) r\Phi^\prime - \left\{ \e^{-2\Psi}\left(1+r\Psi^\prime\right) - 1 + \frac{2q^2}{r^2} \right\} = 0
	\end{equation}
	To simplify this further, we use the substitution, $U = \e^\Phi$, such that, $\Phi^\prime = \frac{U^\prime}{U}$ and $\Phi^\dprime + {\Phi^\prime}^2 = \frac{U^\dprime}{U}$. This leads to,
	\begin{equation}\label{phieq2}
		\e^{-2\Psi}r^2U^\dprime - \e^{-2\Psi}\left(1+r\Psi^\prime\right) rU^\prime - \left\{ \e^{-2\Psi}\left(1+r\Psi^\prime\right) - 1 + \frac{2q^2}{r^2} \right\}U = 0
	\end{equation}
	From equation \eqref{psieq1}, one can get,
	\begin{equation}\label{psiprimeeq}
		\e^{-2\Psi}\left(1+r\Psi^\prime\right) = 1 + m^\prime - \frac{3m}{r} + \frac{\epsilon^\prime}{2} - \frac{3\epsilon}{2r}
	\end{equation} 
	Using equations \eqref{psiprimeeq} and \eqref{psieq1} in equation \eqref{phieq2}, one gets a second order differential equation for $\Phi$. To solve this equation, one needs a specific form of the mass function, $m(r)$, and the charge function, $\epsilon(r)$. In the next section, we do this using the Tolman VII density profile and assuming a constant charge density. However, before that, let us derive the equations used to solve for pressure.
	
	\subsection{Solving for the Pressure}
	Besides the metric coefficients, one also needs the radial dependency of the pressure to analyse the properties of the fluid distribution. A differential equation for pressure can be derived by combining the field equations and the conservation equation. This equation is known as the Tolman-Oppenheimer-Volkoff (TOV) equation \cite{tolman1,opp}. Here, we derive a generalised form of this equation with a charge and a cosmological constant. Using equation \eqref{psieq1} in equation \eqref{efe2}, and separating out $\Phi^\prime$, we get,
	\begin{equation}\label{tovgen1}
		\Phi^{\prime} = \frac{24\pi p r^4 + 6mr + 3\epsilon r - 3r^2\epsilon^\prime + 2\Lambda r^4}{2r^2 \left( 3r - 6m -3\epsilon + \Lambda r^3\right)}  
	\end{equation}  
	Using this in equation \eqref{conseq}, we get the generalised TOV equation,
	\begin{equation}\label{tovgen2}
		p^\prime = -(\rho + p) \frac{24\pi p r^4 + 6mr + 3\epsilon r - 3r^2\epsilon^\prime + 2\Lambda r^4}{2r^2 \left( 3r - 6m -3\epsilon + \Lambda r^3\right)}  + \frac{\epsilon^\dprime}{8\pi r^2} + \frac{\epsilon^\prime}{4\pi r^3}
	\end{equation}
	There is another equation for pressure which may prove more useful in some cases than the TOV equation. This is derived by using, ${G^1}_1 - {G^0}_0 = 8\pi (\rho + p)$, which leads to,
	\begin{equation}\label{press2}
		p = \frac{\e^{-2\Psi}}{4\pi r} \left(\Phi^\prime + \Psi^\prime\right) - \rho
	\end{equation}
	The above equations reinforce the well-known fact that for fluid distributions in GR, given an ansatz for the density profile, the pressure (and hence the equation of state) cannot be arbitrarily chosen but rather must be fixed by solving the field equations.

	\section{The Generalised Tolman VII Solution} \label{sec-tol7}
\begin{figure}[H]
		\centering
		\includegraphics[width=0.5\linewidth]{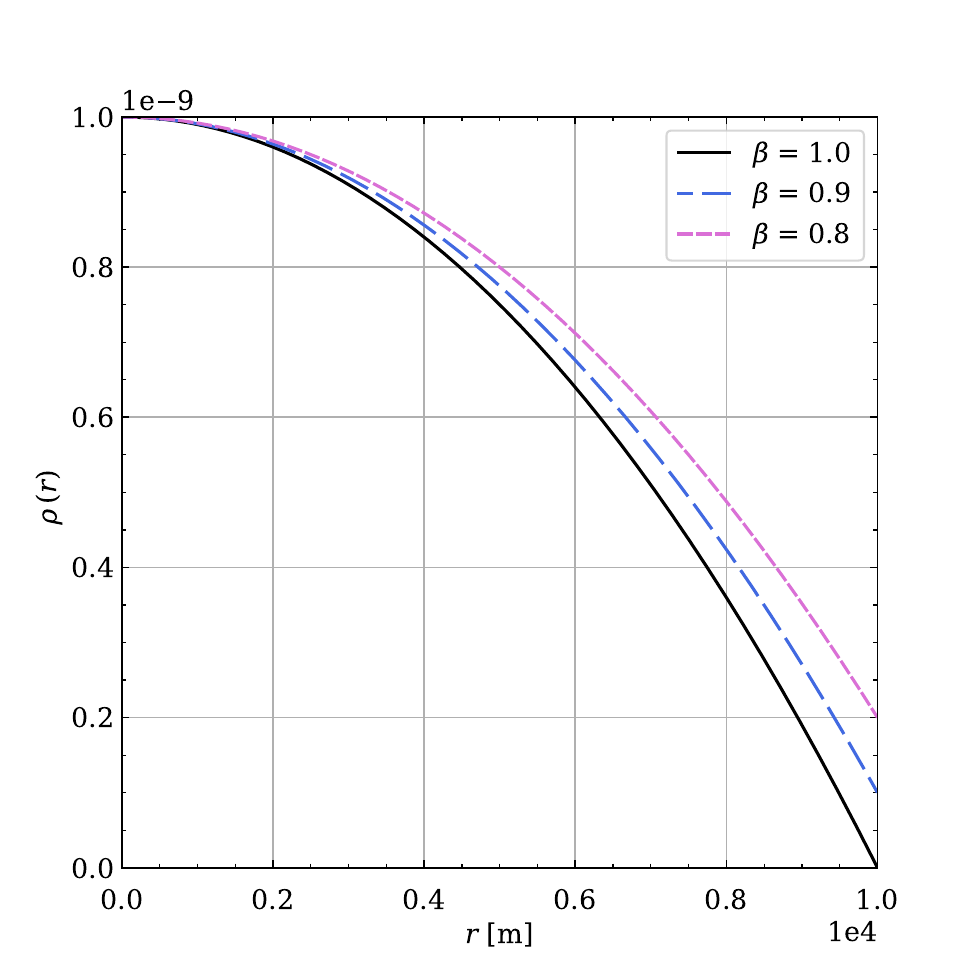}
		\caption{The density profile, $\rho(r)$, as a function of the radius, $r$, for different values of the self-boundedness parameter, $\beta$. The density is continuous and monotonically decreasing ($\rho^\prime < 0$) throughout the extent of the fluid distribution.}\label{fig-den}
	\end{figure}

	The Tolman VII solution is characterised by the following ansatz on the energy density \cite{tolman1,raghoo},
	\begin{equation}\label{rhotol}
		\rho = \rho_0 \left[ 1 - \beta \left( \frac{r}{r_b} \right)^2\right]
	\end{equation} 
	where, $\rho_0$ is the central density ($\rho_0 \equiv\rho(r=0)$), $r_b$ represents a boundary radius beyond which the space-time can be considered to be electro-vacuum, and  $\beta$ is the so-called self-boundedness parameter that takes values between 0 and 1 \cite{raghoo}. It has been shown that for values of $\beta$ other than unity, the solution can be used to model self-bound stars thus extending its physical applicability \cite{raghoo3}.  Using this ansatz, one can solve equation \eqref{phieq2} in an analogous manner as done for the usual Tolman VII solution \cite{mehra,durgapal1}. In figure \ref{fig-den}, we plot the Tolman VII density profile. The density is monotonically decreasing ($\rho^\prime < 0$) for $0<r<r_b$. The values of central density and the boundary radius are given in table \ref{tab-paramval} which correspond to their typical values for a neutron star \cite{raghoo}.
	
	Using equation \eqref{rhotol}, the mass function (equation \eqref{massfunc}) inside the boundary becomes,
	\begin{equation}\label{masstol}
		m(r) = 4\pi \rho_0 \left(\frac{r^3}{3} - \frac{\beta r^5}{5 r_b^2} \right)
	\end{equation}
	Further, we assume a uniform charge density, such that,
	\begin{equation}\label{chargetol}
		q(r) = \frac{Q r^3}{r_b^3} \quad \Rightarrow\ \epsilon(r) = \frac{Q^2r^5}{5r_b^6} 
	\end{equation}
	Using equations \eqref{masstol} and \eqref{chargetol} in equation \eqref{psieq1}, we get\footnote{Originally, in \cite{tolman1}, an ansatz of the form \eqref{psieqtol7} was taken on the metric coefficient and the density profile in equation \eqref{rhotol} was obtained by solving the field equations.},
	\begin{equation}\label{psieqtol7}
		\e^{2\Psi(r)} = \left(1 - \frac{Ar^2}{3} + \frac{Br^4}{5} \right)^{-1}
	\end{equation}
	where, $A := 8\pi \rho_0 - \Lambda$ and $B := \frac{ 8\pi\rho_0  \beta}{r_b^2} - \frac{Q^2}{r_b^6}$. Using the expression for the metric function, equation \eqref{phieq2} becomes,
	\begin{equation}\label{phieqtol1}
		\left( 1 - \frac{Ar^2}{3} + \frac{Br^4}{5} \right) r^2U^\dprime + \left(\frac{Br^4}{5} - 1\right) rU^\prime + \left(\frac{B}{5} - \frac{2Q^2}{r_b^6}\right) r^4 U = 0
	\end{equation}
	Multiplying the above equation by $\frac{5}{B}$; defining a new variable, $x:= r^2$, and writing the derivatives with respect to $x$, the above equation becomes,
	\begin{equation}
		\left( \frac{5}{B} - \frac{5Ax}{3B} + x^2 \right)\tilde{\tilde{U}} + \left(x - \frac{5A}{6B}\right) \tilde{U} + \left(\frac{1}{4} - \frac{5Q^2}{2Br_b^6}\right) U = 0
	\end{equation}
	where, `tilde' represents differentiation with respect to $x$. To simplify this further, we define, $M:=\frac{5}{B} - \frac{5Ax}{3B} + x^2 = \frac{5}{B} \e^{-2\Psi} $; $	N:= x - \frac{5A}{6B} = \frac{\tilde{M}}{2} $; and $K := \frac{1}{4} - \frac{5Q^2}{2Br_b^6} $; such that the above equation takes a simpler form,
	\begin{equation}\label{ueq1}
		M \tilde{\tilde{U}} + N \tilde{U} + K U = 0
	\end{equation}
	To solve this equation, we yet again change all the derivatives to be with respect to a new variable, $w := \ln\left(\sqrt{M}+N\right)$. This gives, $\tilde{w}=\frac{1}{\sqrt{M}}$, and, $ \tilde{\tilde{w}} = -\frac{N}{M\sqrt{M}} $. Using this, we get, $\tilde{U} = \frac{1}{\sqrt{M}}\frac{\df U}{\df w}$, and, $ \widetilde{\widetilde{U}} = \frac{1}{M}\frac{\df^2 U}{\df w^2} - \frac{N}{M\sqrt{M}}\frac{\df U}{\df w}$. Substituting these in equation \eqref{ueq1}, we get,
	\begin{equation}
		\frac{\df ^2U}{\df w^2} + K U = 0
	\end{equation}
	
	The solution to the above equation is given by\footnote{One can, in fact, work with a less general solution comprising of only either a (hyperbolic) cosine or a (hyperbolic) sine term as done for the uncharged case in \cite{stuch2}. This reduces the complexity of the calculations since there would be only one integration constant to fix. However, here, we choose to work with the more general form of the solution.},
	\begin{equation}\label{phieq3}
		U(w)= \e^\Phi = f_K(w) = 
		\begin{cases}
			C_1 \cos(Dw) + C_2 \sin(Dw)  &\qquad {\rm if}\ K > 0\\
			C_1 w + C_2 &\qquad {\rm if}\ K = 0 \\
			C_1 \cosh(Dw) + C_2 \sinh(Dw) &\qquad {\rm if}\ K < 0
		\end{cases}
	\end{equation}
	where, to make the notation less tedious, we have introduced the variable, $D:= \sqrt{|K|}$, and $C_1$, $C_2$ are arbitrary integration constants. In terms of the fundamental parameters of the system, the three conditions on $K$ correspond to $8\pi\rho_0\beta r_b^4 - 11Q^2 >,=,< 0 $. In the absence of charge, the latter two cases do not exist since $K = \frac{1}{4} > 0$. Therefore, it is the inclusion of charge in the Tolman VII space-time that is leading to these three distinct subclasses of solutions.
	
\begin{figure}[H]
     \centering
     \includegraphics[width=\linewidth]{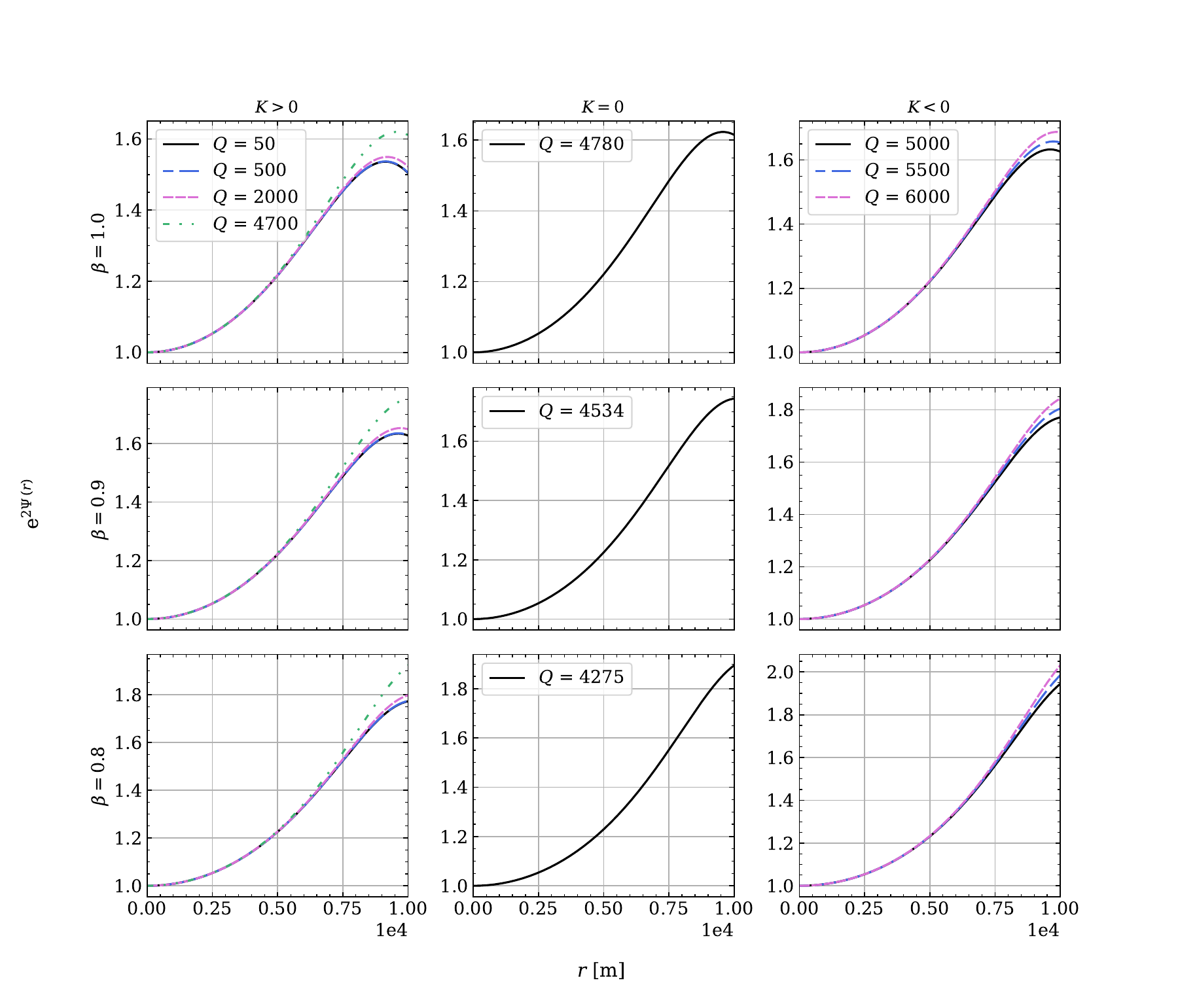}
     \caption{The metric function, $\e^{2\Psi}$ plotted as a function of radial coordinate, $r$, in the three different subclasses of the solution. The values of various parameters are listed in table \ref{tab-paramval}.}
     \label{fig-e2psi}
 \end{figure}

	Since in each case, there is a distinct constraint on the constant $B$, the metric function in equation \eqref{psieq1} needs to be analysed separately for $K>,=,<0$. In figure \ref{fig-e2psi}, we plot the metric function as a function of the radial coordinate for each case. The values of the various parameters corresponding to each case are listed in table \ref{tab-paramval}.
	
	\begin{table}[H]
		\centering
		\begin{tabular}{ c  c  c  c  c  c } 
			\hline 
			\hline
			
			$\rho_0 $ & $r_b $ & $\Lambda$ & & $Q/\beta^\hf$ &   \\ [0.1ex]
			& & & $K>0$ & $K = 0$ & $K<0$\\	[0.1ex]
			\hline  
			& & & & &\\ [-2.0ex]
			$ 10^{-9}$ m$^{-2}$ & $10^{4}$ m & $10^{-52}$ m$^{-2}$ & $<4780 $ m & $ 4780 $ m & $>4780 $ m\\   [0.1ex]
			$ 10^{18}$ kg m$^{-3}$ & $10^{4}$ m & $10^{-52}$ m$^{-2}$ & -- & $5.56 \times 10^{20}$ C & --\\ 
			\hline
			\hline
		\end{tabular}
		\caption{The orders of magnitude of various parameters in geometric and standard units. The value of $\Lambda$ is fixed via cosmology \cite{planck} while the values of other constants correspond to their typical values for compact objects such as neutron star or self-bound stars \cite{raghoo,ray}.}
		\label{tab-paramval}
	\end{table}

	Besides the metric functions, we also need the expression for the pressure of the fluid. For this, we take a derivative of equation \eqref{phieq3} with respect to $r$ to get,
	\begin{equation}\label{phipri}
		\Phi^\prime = \frac{U^\prime}{U} = 2rD\sqrt{\frac{B}{5}}\e^{\Psi-\Phi}\ h_K(w)
	\end{equation}
	where,
	\begin{equation}
		h_K(w) = \frac{1}{D}\tilde{f}_K(w) = \begin{cases}
			C_2 \cos(Dw) - C_1 \sin(Dw)  &\qquad {\rm if}\ K > 0\\
			\frac{C_1}{D}  &\qquad {\rm if}\ K = 0 \\
			C_2 \cosh(Dw) + C_1 \sinh(Dw) &\qquad {\rm if}\ K < 0
		\end{cases}
	\end{equation}
	Using equation \eqref{phipri} in equation \eqref{press2}, we get,
	\begin{equation}\label{press3}
		p = \frac{D}{2\pi}\sqrt{\frac{B}{5}}\ \e^{-\Psi} \ \frac{h_K(w)}{f_K(w)} - \frac{1}{2\pi}\left(\frac{B}{5}r^2 - \frac{A}{6}\right)  - \rho
	\end{equation}
	To find the complete expressions for the metric function, $\e^{2\Phi}$, and the pressure, one needs to fix the constants, $C_1$ and $C_2$. Before doing that, we present some features of the spatial geometry of the solution since it does not involve the integration constants.
	
	\subsection{Spatial Geometry}
	The line element in equation \eqref{ssle} can be written as,
	\begin{equation}
		\df s^2 = - \e^{2\Phi}\df t^2 + \frac{\df r^2}{1 - k(r)r^2} + r^2 \df \theta^2 + r^2 \sin^2 \theta \df \phi^2
	\end{equation}
	where,
	\begin{equation}
		k(r):= \frac{2m(r)}{r^3} + \frac{\epsilon(r)}{r^3} - \frac{\Lambda}{3} = \frac{^{(3)}R}{6}
	\end{equation}
	where, $^{(3)}R$ is the Ricci scalar for the $t = {\rm const.}$ (spatial) hypersurfaces. The function, $k(r)$,  determines the spatial curvature. Depending on whether this function is positive, zero, or negative, we will have a fluid sphere with spherical, flat, and hyperbolic spatial geometry, respectively. Using equations \eqref{masstol} and \eqref{chargetol}, for the generalised Tolman VII space-time, we get,
	\begin{equation}
		k(r) = \frac{A}{3} - \frac{B}{5}r^2 = \frac{1}{3}\left(8\pi \rho_0 - \Lambda\right) - \left(\frac{ 8\pi\rho_0  \beta}{r_b^2} - \frac{Q^2}{r_b^6}\right)\frac{r^2}{5} 
	\end{equation}

    \subsection*{Constant Spatial Curvature}
	For $8\pi\rho_0\beta r_b^4 = Q^2$, we have a constant spatial curvature that depends only on the variable, $A$. This is only a possibility in the subclass with $K < 0$ since $K\ge 0$ means $8\pi \rho_0 \beta r_b^4 \ge 11Q^2$. The positive, zero, or negative curvature will then corresponds to, $8\pi\rho_0 >,=,< \Lambda$, respectively. Curiously, these conditions are the same as the ones for a uniform density fluid sphere \cite{bohm3}. Below, we derive general conditions on $\rho_0,\ \Lambda,\ Q$, and $r_b$ for the three types of the spatial curvature.
	
	\subsubsection*{Spherical Spatial Geometry}
	For the spatial geometry to be spherical, we need the spatial curvature, $k(r)$, to be positive. This gives,
	\begin{equation} 
		\frac{A}{3} > \frac{Br^2}{5} \qquad \Rightarrow\  r^2 < \frac{5A}{3B}\label{poscurvcond1} 
	\end{equation}

	Therefore, depending on the values of the constants, $\rho_0$, $\Lambda$, and $Q^2$, we will get the range for the radial coordinate up to which, the fluid sphere will have a spherical spatial geometry. If the right hand side of the inequality \eqref{poscurvcond1} is greater than $r_b^2$, then we get a space-time that has positively curved spatial hypersurfaces for the whole extent of the fluid distribution. This condition takes the form,
	\begin{equation} \label{poscurvcond3}
		5A > 3Br_b^2 \qquad \Rightarrow\ 8\pi\rho_0(5-3\beta) > 5\Lambda - \frac{3Q^2}{r_b^4}
	\end{equation}
	For $K>0$, we have, $Br_b^2 > \frac{10Q^2}{r_b^4}$. Therefore, in this case, a necessary (but not sufficient) condition for having a positively curved spatial geometry throughout becomes,
	\begin{equation}\label{poscurvcond4}
		8\pi\rho_0 > \Lambda + \frac{6Q^2}{r_b^4}
	\end{equation}
	For $K = 0$, the above condition is necessary and sufficient while for $K < 0$ it is sufficient (but not necessary).
	
	\subsubsection*{Flat Spatial Geometry}
	The spatial curvature is a continuous function of $r$. Therefore, unlike the other two cases, the spatial geometry can be flat only at a given point, not throughout the fluid distribution. This is clear from the condition for flat spatial geometry which is obtained by replacing the inequality sign in equation \eqref{poscurvcond1} by an equality,
	\begin{equation}
		r^2 = \frac{5A}{3B}
	\end{equation} 
	If indeed there exists such a point, then this point will lie inside the boundary of the fluid sphere if $5A < 3Br_b^2$. For $K>0$ a necessary condition for this to happen is the same as equation \eqref{poscurvcond4}. For $K = 0$, it becomes a necessary and sufficient condition while for $K<0$, it is a sufficient condition. 
	
	\subsubsection*{Hyperbolic Spatial Geometry}
	For the spatial curvature to be negative, the condition in equation \eqref{poscurvcond1} will be reversed, and we get,
	\begin{equation}
		r^2 > \frac{5A}{3B} 
	\end{equation}
	If one has either $A<0$ and $ B>0 $ or $A>0$ and $B<0$, then the above condition is satisfied for all values of the radial coordinate. For $K>0$ and $K=0$, we have, $B>0$. Then, for the spatial geometry to be hyperbolic, it is necessary and sufficient to have, $8\pi\rho_0 < \Lambda$. For $K<0$, we have two cases: (i) if $B>0$, then we need, $\Lambda > 8\pi\rho_0$. This will be satisfied if we have, $\Lambda > \frac{11Q^2}{\beta r_b^4}$. (ii) if $B<0$, then we need, $\Lambda < 8\pi \rho_0$. A necessary condition for this is, $\Lambda < \frac{11Q^2}{\beta r_b^4}$.
	
\begin{figure}[H]
        \centering
        \includegraphics[width=\linewidth]{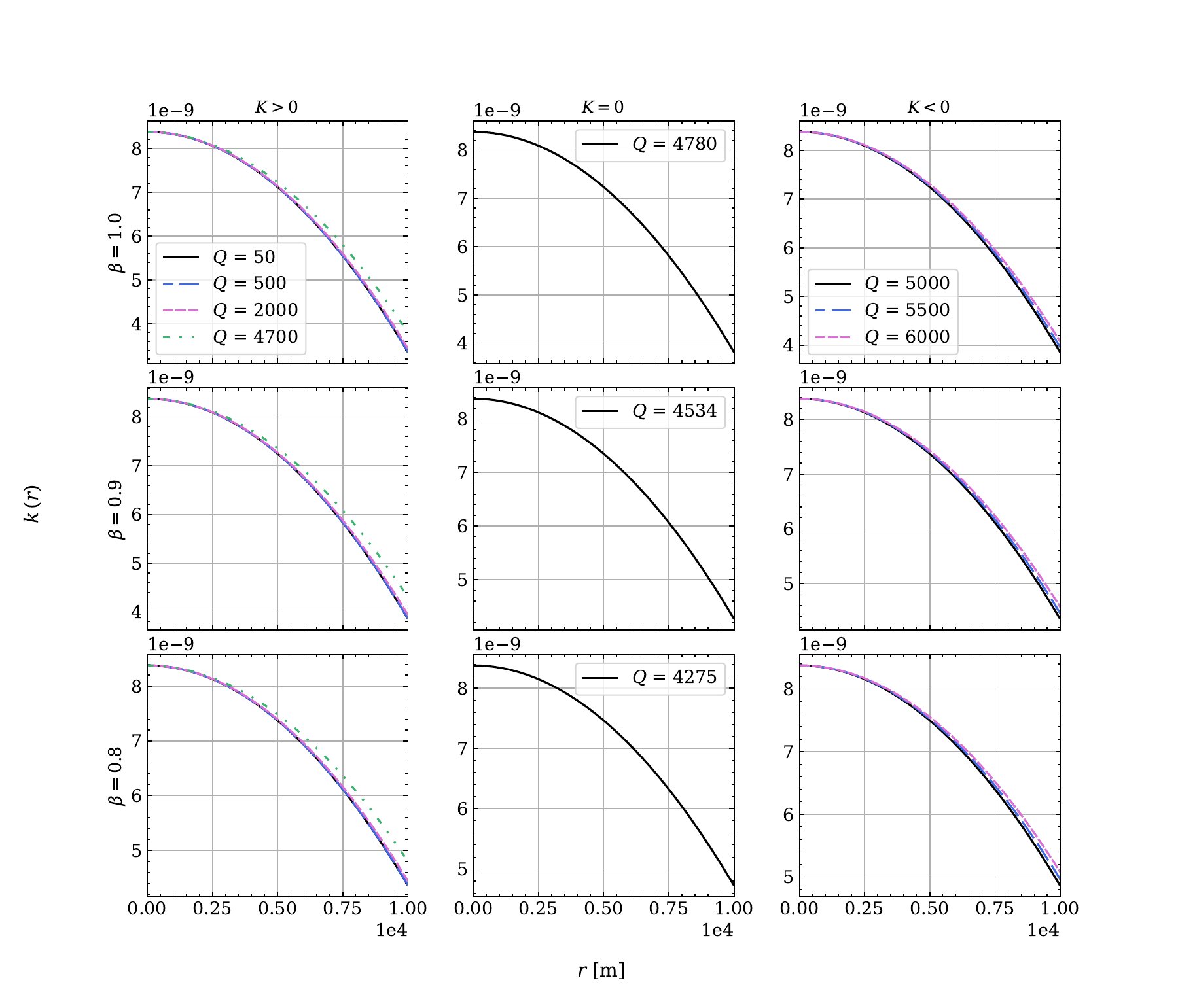}
        \caption{The spatial curvature, $k(r)$, plotted as a function of the radial coordinate, $r$, for the three subclasses of the solution. The values of various parameters are listed in table \ref{tab-paramval}. For these values, all three subclasses exhibit a positive spatial curvature (spherical spatial geometry).}
        \label{fig-spatialcurv}
    \end{figure}

	The analysis in this section has shown that it is possible to have all three types of spatial geometries inside the fluid sphere. We plot the spatial curvature for the three different subclasses of solution ($K>,=,<0$) in figure \ref{fig-spatialcurv}. 
	
	\section{Matching with the Exterior Space-time} \label{sec-extst}
	
	We are analysing a fluid distribution that has a finite radial extent, $r \le r_b$, where $r_b$ is the boundary radius. Beyond this ($r\ge r_b$), one can take the density and pressure to vanish $(\rho = 0 = p)$. In the presence of a charge and a cosmological constant, this exterior region can be described by either of the two well-known solutions -- the Reissner-Nordstr\"om de Sitter space-time \cite{stephani} and the charged Nariai space-time \cite{nari1,nari2,bohm2}. The line elements for these two space-times are, respectively, given by,
	\begin{subequations}
		\begin{equation}\label{rsds}
			\df s^2 = -\left(1 - \frac{2M}{r} + \frac{Q^2}{r^2} + \frac{\Lambda r^2}{3}\right)\df t^2 + \frac{\df r^2}{\left(1 - \frac{2M}{r} + \frac{Q^2}{r^2} + \frac{\Lambda r^2}{3}\right)} + r^2 \left( \df \theta^2 + \sin^2\theta\df\phi^2 \right)
		\end{equation}
		\begin{equation}\label{nariai}
			\df s^2 = \frac{1}{\Lambda - \frac{Q^2}{r_b^4}}\left[ -\left(X\cos u + Y\sin u \right)^2 \df t^2 + \df u^2 \right] + \frac{1}{\Lambda + \frac{Q^2}{r_b^4}} \left( \df \theta^2 + \sin^2\theta\df\phi^2 \right)
		\end{equation}
		where, $M = m(r_b) + \frac{\epsilon(r_b)}{2} + \frac{Q^2}{2r_b}$ and $Q = q(r_b)$ are, respectively, the total mass\footnote{This is the total gravitational mass, $M = m_g(r_b)$, where, the gravitational mass is defined as \cite{bek,dev},
			\begin{equation}
				m_g (r) := m(r) + \frac{\epsilon(r)}{2} + \frac{q^2(r)}{2r}
		\end{equation} } and the total charge contained within the region, $r\le r_b$; $X$ and $Y$ are arbitrary constants. In what follows, we present the steps necessary to match the Tolman VII solution to both these exterior space-times.
	\end{subequations}

	\subsection{Solutions Interior to Reissner-Nordstr\"om-de Sitter Space-time}
	To get the full expression for the metric coefficients and the pressure, we need to find the values of the constants in equations \eqref{phieq3} and \eqref{press3}. This can be done by matching the interior solution with the Reissner-Nordstr\"om-de Sitter space-time at the boundary. It is straightforward to see that, $\e^{2\Psi(r_b)}|_{\rm TVII} = \e^{2\Psi(r_b)}|_{\rm RNdS}$. Additionally, we will also demand that at the boundary radius, $r_b$, the pressure vanishes ($p(r_b) = 0$). This, along with, $\e^{-\Psi(r_b)} = \e^{\Phi(r_b)} = f_K(w_b) $, used in equations \eqref{phieq3} and \eqref{press3}, leads to,
	\begin{subequations}
		\begin{align}
			f_K(w_b) &= \left(1 - \frac{Ar_b^2}{3} + \frac{Br_b^4}{5}\right)^{\hf}\\
			h_K(w_b) &= \frac{1}{D}\sqrt{\frac{5}{B}}\left\{\frac{B}{5}r_b^2 - \frac{A}{6} + 2\pi \rho_0(1-\beta) \right\}
		\end{align}
	\end{subequations}
	where, $w_b = w(r_b)$. The above two equations can be solved easily to get,
	\begin{subequations} \label{tolconst}
		\begin{align}
			C_1 &=
			\begin{cases}
				\scalemath{0.8}{\left(1 - \frac{Ar_b^2}{3} + \frac{Br_b^4}{5}\right)^{\hf} \cos(Dw_b) - \frac{1}{D}\sqrt{\frac{5}{B}}\left\{\frac{B}{5}r_b^2 - \frac{A}{6} + 2\pi \rho_0(1-\beta) \right\} \sin(Dw_b)} &\quad {\rm if}\ K > 0\\
				\sqrt{\frac{5}{B}}\left\{\frac{B}{5}r_b^2 - \frac{A}{6} + 2\pi \rho_0(1-\beta) \right\} &\quad {\rm if}\ K = 0\\
				\scalemath{0.8}{\left(1 - \frac{Ar_b^2}{3} + \frac{Br_b^4}{5}\right)^{\hf} \cosh(Dw_b) - \frac{1}{D}\sqrt{\frac{5}{B}}\left\{\frac{B}{5}r_b^2 - \frac{A}{6} + 2\pi \rho_0(1-\beta) \right\} \sinh(Dw_b)} &\quad {\rm if}\ K < 0				
			\end{cases}\\
			C_2 &= 
			\begin{cases}
				\scalemath{0.8}{\left(1 - \frac{Ar_b^2}{3} + \frac{Br_b^4}{5}\right)^{\hf} \sin(Dw_b) + \frac{1}{D}\sqrt{\frac{5}{B}}\left\{\frac{B}{5}r_b^2 - \frac{A}{6} + 2\pi \rho_0(1-\beta) \right\} \cos(Dw_b)} &\quad {\rm if}\ K > 0\\
				\left(1 - \frac{Ar_b^2}{3} + \frac{Br_b^4}{5}\right)^{\hf} - \sqrt{\frac{5}{B}}\left\{\frac{B}{5}r_b^2 - \frac{A}{6} + 2\pi \rho_0(1-\beta) \right\}w_b &\quad {\rm if}\ K = 0\\
				\scalemath{0.8}{\frac{1}{D}\sqrt{\frac{5}{B}}\left\{\frac{B}{5}r_b^2 - \frac{A}{6} + 2\pi \rho_0(1-\beta) \right\} \cosh(Dw_b) - \left(1 - \frac{Ar_b^2}{3} + \frac{Br_b^4}{5}\right)^{\hf} \sinh(Dw_b)} &\quad {\rm if}\ K < 0
			\end{cases}
		\end{align}
	\end{subequations}

\begin{figure}[H]
            \centering
            \includegraphics[width=\linewidth]{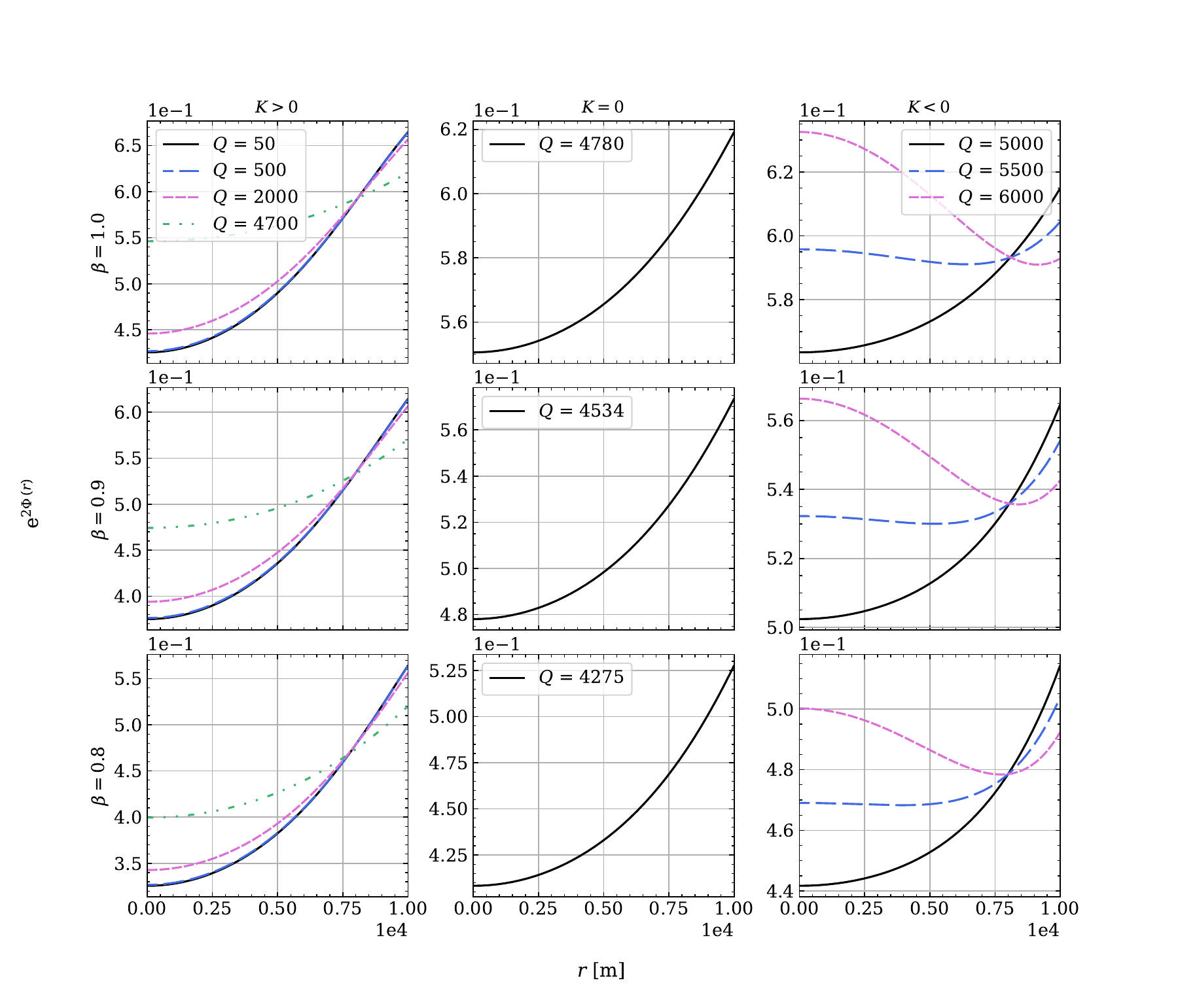}
            \caption{The metric function, $\e^{2\Phi}$, plotted as a function of the radial coordinate, $r$, for the three subclasses of the solution. The values of various parameters are listed in table \ref{tab-paramval}. For all three cases, the qualitative behaviour of the metric function remains the same up to $Q \sim 5500$ m.}
            \label{fig-e2phi}
        \end{figure}

	Using these in equation \eqref{phieq3}, we get the following expression for the metric coefficient,
	\begin{equation}\label{phieqtol7}
		\e^{2\Phi} =
		\begin{cases}
			\scalemath{0.7}{\left[\left(1 - \frac{Ar_b^2}{3} + \frac{Br_b^4}{5}\right)^{\hf} \cos[D(w-w_b)] + \frac{1}{D}\sqrt{\frac{5}{B}}\left\{\frac{B}{5}r_b^2 - \frac{A}{6} + 2\pi \rho_0(1-\beta) \right\} \sin[D(w-w_b)]\right]^2} &\quad {\rm if}\ K > 0 \\
			\scalemath{0.8}{\left[\left(1 - \frac{Ar_b^2}{3} + \frac{Br_b^4}{5}\right)^{\hf} + \sqrt{\frac{5}{B}}\left\{\frac{B}{5}r_b^2 - \frac{A}{6} + 2\pi \rho_0(1-\beta) \right\}(w-w_b)\right]^2} &\quad {\rm if}\ K = 0\\
			\scalemath{0.7}{\left[\left(1 - \frac{Ar_b^2}{3} + \frac{Br_b^4}{5}\right)^{\hf} \cosh[D(w-w_b)] + \frac{1}{D}\sqrt{\frac{5}{B}}\left\{\frac{B}{5}r_b^2 - \frac{A}{6} + 2\pi \rho_0(1-\beta) \right\} \sinh[D(w-w_b)]\right]^2} &\quad {\rm if}\ K < 0
		\end{cases}
	\end{equation}
	In figure \ref{fig-e2phi}, we plot the metric function above as a function of the radial coordinate. 
	
	\subsection{Solutions Interior to the Charged Nariai Space-time}
	In the previous section, we calculated the value of the integration constants by matching the Tolman VII solution to the Reissner-Nordstr\"om (anti-)de Sitter solution. Now, our goal is to match it to the charged Nariai solution. However, the charged Nariai solution (equation \eqref{nariai}) itself has arbitrary constants, $X$ and $Y$, that need to be fixed. Therefore, we will not be able to determine the Tolman VII constants, $C_1$ and $C_2$, by matching the metric coefficients alone. To overcome this problem, we introduce a new quantity -- the central pressure, $p_0 = p(0)$. This along with, $p(r_b) = 0$, gives us,
	\begin{subequations}
		\begin{align}
			h_K(w_0) &= \frac{1}{D}\sqrt{\frac{5}{B}}\left\{2\pi (\rho_0 + p_0) - \frac{A}{6}  \right\} \\
			h_K(w_b) &= \frac{1}{D}\sqrt{\frac{5}{B}}\left\{\frac{B}{5}r_b^2 - \frac{A}{6} + 2\pi \rho_0(1-\beta) \right\}
		\end{align}
	\end{subequations}
	where, $w_0 = w(0)$ and $w_b = w(r_b)$. Using these, we can solve for the constants to get,
	\begin{subequations}\label{tolconst1}
		\begin{align}
			C_1 &= 
			\begin{cases}
				\frac{\frac{1}{D}\sqrt{\frac{5}{B}}\left[ \left\{2\pi (\rho_0 + p_0) - \frac{A}{6}  \right\} \cos(Dw_b) - \left\{\frac{B}{5}r_b^2 - \frac{A}{6} + 2\pi \rho_0(1-\beta) \right\} \cos(Dw_0) \right]}{\sin[D(w_b - w_0)]} \quad &{\rm if}\ K > 0 \\
				\sqrt{\frac{5}{B}}\left\{2\pi (\rho_0 + p_0) - \frac{A}{6}  \right\} \quad &{\rm if}\ K = 0 \\
				-\frac{\frac{1}{D}\sqrt{\frac{5}{B}}\left[ \left\{2\pi (\rho_0 + p_0) - \frac{A}{6}  \right\} \cosh(Dw_b) - \left\{\frac{B}{5}r_b^2 - \frac{A}{6} + 2\pi \rho_0(1-\beta) \right\} \cosh(Dw_0) \right]}{\sinh[D(w_b - w_0)]} \quad &{\rm if}\ K < 0
			\end{cases}\\
			C_2 &= 
			\begin{cases}
				\frac{\frac{1}{D}\sqrt{\frac{5}{B}}\left[ \left\{2\pi (\rho_0 + p_0) - \frac{A}{6}  \right\} \sin(Dw_b) - \left\{\frac{B}{5}r_b^2 - \frac{A}{6} + 2\pi \rho_0(1-\beta) \right\} \sin(Dw_0) \right]}{\sin[D(w_b - w_0)]} \quad &{\rm if}\ K > 0 \\
				{\rm undetermined} \quad &{\rm if}\ K = 0 \\
				\frac{\frac{1}{D}\sqrt{\frac{5}{B}}\left[ \left\{2\pi (\rho_0 + p_0) - \frac{A}{6}  \right\} \sinh(Dw_b) - \left\{\frac{B}{5}r_b^2 - \frac{A}{6} + 2\pi \rho_0(1-\beta) \right\} \sinh(Dw_0) \right]}{\sinh[D(w_b - w_0)]} \quad &{\rm if}\ K < 0
			\end{cases}
		\end{align}
	\end{subequations}
	The above values of the constants are valid when matching with the Reissner-Nordstr\"om de Sitter solution as well. However, with this treatment, one more parameter -- the central pressure -- would get added to the already complicated parameter space of the Tolman VII solution. Hence we avoided doing this in the previous section where it was not necessary.
	
	We note that the metric functions in the Nariai solution contain trigonometric functions and so does the Tolman VII solution for $K>0$. Further, $C_2$ remains undetermined for the case with $K = 0$ while the $K<0$ case contains hyperbolic trigonometric functions. Considering this, the case with $K>0$ is the simplest candidate to be an interior to the Nariai solution\footnote{It is possible to use a similar treatment as in this section for the $K = 0$ case as well.}. The line element for the Tolman VII solution in this case is given by,
	\begin{equation}
		\df s^2 = - \left[C_1 \cos(Dw) + C_2 \sin(Dw)\right]^2 \df t^2 + \e^{2\Psi} \df r^2 + r^2\df\theta^2 + r^2\sin^2 \theta\df \phi^2
	\end{equation}
	where, $C_1$ and $C_2$ are now given by equations \eqref{tolconst1}. Now, using the fact that, $\frac{\df w}{\df x} = \frac{1}{\sqrt{M}}$ and $\frac{1}{M} = \frac{B\e^{2\Psi}}{5}$, we get, 
	\begin{equation}
		\e^{2\Psi}\df r^2 = \frac{5}{4B} \frac{\df w^2}{r^2}
	\end{equation}
	Using this, the line element becomes,
	\begin{equation}
		\df s^2 = - \left[C_1 \cos(Dw) + C_2 \sin(Dw)\right]^2 \df t^2 + \frac{5}{4B} \frac{\df w^2}{r^2} + r^2\df\theta^2 + r^2\sin^2 \theta\df \phi^2
	\end{equation}
	Following the treatment in \cite{bohm,bohm2}, we introduce a new coordinate, $ u = Dw$, to get,
	\begin{equation}
		\df s^2 = - \left[C_1 \cos u + C_2 \sin u\right]^2 \df t^2  + \frac{5}{4BD^2} \frac{\df u^2}{r^2} + r^2\df\theta^2 + r^2\sin^2 \theta\df \phi^2
	\end{equation}
	Comparing the above line element to that in equation \eqref{nariai}, we see that the metric functions corresponding to the $\df t^2$ term will be identical for the two line elements if the constants, $X$ and $Y$, are given by,
	\begin{align}
		X &= C_1\sqrt{\Lambda  - \frac{Q^2}{r_b^4}} &=  \scalemath{0.75}{\frac{\frac{1}{D}\sqrt{\frac{5}{B}} \sqrt{\Lambda  - \frac{Q^2}{r_b^4}}\left[ \left\{2\pi (\rho_0 + p_0) - \frac{A}{6}  \right\} \cos(Dw_b) - \left\{\frac{B}{5}r_b^2 - \frac{A}{6} + 2\pi \rho_0(1-\beta) \right\} \cos(Dw_0) \right]}{\sin[D(w_b - w_0)]}} \\
		Y &= C_2\sqrt{\Lambda  - \frac{Q^2}{r_b^4}} &= \scalemath{0.75}{\frac{\frac{1}{D}\sqrt{\frac{5}{B}} \sqrt{\Lambda  - \frac{Q^2}{r_b^4}}\left[ \left\{2\pi (\rho_0 + p_0) - \frac{A}{6}  \right\} \sin(Dw_b) - \left\{\frac{B}{5}r_b^2 - \frac{A}{6} + 2\pi \rho_0(1-\beta) \right\} \sin(Dw_0) \right]}{\sin[D(w_b - w_0)]}}
	\end{align}
	
	Once $X$ and $Y$ are fixed in this way, the two space-times can be matched by requiring that at the boundary radius, $r_b$, we have,
	\begin{equation}
		\frac{5}{4BD^2r_b^2} = \frac{1}{\Lambda  - \frac{Q^2}{r_b^4}} \quad {\rm and} \quad r_b^2 = \frac{1}{\Lambda  + \frac{Q^2}{r_b^4}}
	\end{equation}
	Using, $4BD^2r_b^2 = 8\pi\rho_0\beta - \frac{11Q^2}{r_b^4}$, the conditions above translate to the following,
	\begin{equation}
		\frac{6Q^2}{r_b^4} = 8\pi\rho_0\beta - 5\Lambda \quad {\rm and} \quad \frac{Q^2}{r_b^4} = \frac{1}{r_b^2} - \Lambda
	\end{equation}
	Subtracting the second condition from the first, we get,
	\begin{equation}
		(8\pi\rho_0\beta - 4\Lambda) r_b^4 - r_b^2 - 5Q^2 = 0
	\end{equation}
	This can be easily solved to get,
	\begin{equation} \label{narimatch}
		r_b^2 = \frac{1 \pm \sqrt{1 + 20 Q^2(8\pi\rho_0\beta - 4\Lambda)}}{2(8\pi\rho_0\beta - 4\Lambda)}
	\end{equation}
	For realistic values of the central density and the cosmological constant, the quantity in the square root above is greater than one. Hence, requiring the boundary radius to be real, we can ignore the negative sign in the numerator. For realistic values, $\Lambda = 10^{-52}$ m$^{-2}$, $Q = 10^3$ m, and $\rho_0 = 10^{-9}$ m$^{-2}$, the boundary radius turns out to be, $r_b \sim 6500$ m, which is a reasonable size for a compact star \cite{raghoo}. 	
	
    \section{Trapping of Null Geodesics} \label{sec-geodestrap}
	The Tolman VII solution has been shown to exhibit internal trapping of null geodesics \cite{ishak,neary,stuch2}. Study of geodesic trapping provides an insight into the geodesic structure of the space-time and is also relevant in astrophysics due to the possibility of trapping of gravitational waves and neutrinos \cite{stuch2}. Here, we investigate the effects of the self-boundedness parameter, cosmological constant, and charge on the trapping of null geodesics.
	
	\subsection{Effective Potential}	
	Due to spherical symmetry, we can fix, $\theta = \frac{\pi}{2}$ (equatorial plane). Then, the geodesic equations for the line element \eqref{ssmetric} take the form,
	\begin{align}
		\ddot{t} + 2\dot{r}\dot{t}\Phi^\prime &= 0 \label{geoeq1}\\
		\ddot{r} + \dot{r}^2{\Psi^\prime} + e^{2\Phi - 2\Psi}\dot{t}^2{\Phi^\prime} - \e^{-2\Psi} r \dot{\phi}^2 &= 0 \label{geoeq2}\\
		\ddot{\phi} + \frac{2\dot{r}\dot{\phi}}{r} &= 0 \label{geoeq3}
	\end{align}
	Additionally, for null geodesics, we also have, 
	\begin{equation}\label{nullgeo}
		-\e^{2\Phi} \dot{t}^2 + \e^{2\Psi} \dot{r}^2 + r^2 \dot{\phi}^2 = 0
	\end{equation}
	Here, `dot' represents a derivative with respect of some affine parameter. The first integrals of equations \eqref{geoeq1} and \eqref{geoeq3} lead to the familiar conserved quantities,
	\begin{equation}\label{consquant}
		\e^{2\Phi}\dot{t} = E \quad {\rm and} \quad r^2\dot{\phi} = \ell
	\end{equation}
	Using this in equation \eqref{nullgeo} leads to,
	\begin{equation}
		r^2\dot{r}^2 = \e^{-2\Psi}E^2 \left(r^2 \e^{-2\Phi} - \frac{\ell^2}{E^2}\right)
	\end{equation}
	We define the impact parameter as, $b = \frac{\ell}{E}$ \cite{mtw,ishak}. Then, one can get the turning points of the geodesic motion by introducing an effective potential, $V_{\rm eff} (r) = r^2 \e^{-2\Phi}$, such that, the above equation leads to the constraint, $b^2 \le V_{\rm eff}$. This is an invariant definition of the effective potential valid for all static perfect fluid spheres \cite{ishak}.
	
	For the Tolman VII space-time, we use equation \eqref{phieqtol7} to get the following forms of the effective potential for the three different classes of solution,
	\begin{equation}\label{vefftol7}
		V_{\rm eff}(r) = 
		\begin{cases}
			\left[\frac{r}{\scalemath{0.7}{\left(1 - \frac{Ar_b^2}{3} + \frac{Br_b^4}{5}\right)^{\hf} \cos[D(w-w_b)] + \frac{1}{D}\sqrt{\frac{5}{B}}\left\{\frac{B}{5}r_b^2 - \frac{A}{6} + 2\pi \rho_0(1-\beta) \right\} \sin[D(w-w_b)]}}\right]^2 &\quad {\rm if}\ K>0\\
			\scalemath{0.9}{\left[\frac{r}{\left(1 - \frac{Ar_b^2}{3} + \frac{Br_b^4}{5}\right)^{\hf} + \sqrt{\frac{5}{B}}\left\{\frac{B}{5}r_b^2 - \frac{A}{6} + 2\pi \rho_0(1-\beta) \right\}(w-w_b)}\right]^2} &\quad {\rm if}\ K = 0\\
			\left[\frac{r}{\scalemath{0.68}{\left(1 - \frac{Ar_b^2}{3} + \frac{Br_b^4}{5}\right)^{\hf} \cosh[D(w-w_b)] + \frac{1}{D}\sqrt{\frac{5}{B}}\left\{\frac{B}{5}r_b^2 - \frac{A}{6} + 2\pi \rho_0(1-\beta) \right\} \sinh[D(w-w_b)]}}\right]^2 &\quad {\rm if}\ K < 0
		\end{cases}
	\end{equation}
	We will have a circular null orbit of radius, $r_c$, if, $\left.\frac{\df V_{\rm eff}}{\df r}\right|_{r_c} = 0$. If we have, $r_c < r_b$, then the circular orbits will be trapped inside the fluid sphere.
	
%

	\subsection{Conditions for Trapped Null Orbits}
	For having trapped null orbits, we require that for some $r_c < r_b$, we have,
	\begin{equation}\label{trapnullcond0}
		\left.\frac{\df V_{\rm eff}}{\df r}\right|_{r_c} = 0
	\end{equation}
	Using, $V_{\rm eff} (r) = r^2 \e^{-2\Phi}$, this reduces to,
	\begin{equation}\label{trapnullcond}
		r_c\Phi^\prime(r_c) = 1 
	\end{equation}
	This is a necessary and sufficient condition for the existence of trapped null orbits\footnote{An alternate way to derive this condition (without using $V_{\rm eff}$) is through equation \eqref{geoeq2}. Using equation \eqref{nullgeo} in \eqref{geoeq2}, we get,
		\begin{equation}
			\ddot{r} + \dot{r}^2 \left(\Psi^\prime + \Phi^\prime\right) + \e^{-2\Psi} \dot{\phi}^2 r \left(r\Phi^\prime - 1 \right) = 0
		\end{equation}
		The space-time will have a trapped circular orbit if at some, $r_c<r_b$, we have, $\dot{r} = 0 = \ddot{r}$. This reduces the above equation to, $r\Phi^\prime = 1$, which is the same condition as equation \eqref{trapnullcond}.
	}. The geometric condition above can also be converted into a condition on the physical properties of the fluid. To do this, we use equation \eqref{tovgen1}, which gives us,
	\begin{equation}\label{trapnullcond1}
		4\pi p r^2 + \frac{3m}{r} + \frac{3}{2}\frac{\epsilon}{r} - \hf \epsilon^\prime = 1
	\end{equation}
	It is interesting to note that this condition does not explicitly contain the cosmological constant. Further, using equation \eqref{conseq} or the TOV equation \eqref{tovgen2}, we can also write\footnote{Similarly, conditions involving both geometric and physical quantities can also be derived. For example, combining equation \eqref{trapnullcond1} with \eqref{psiprimeeq}, we get,
		\begin{equation}
			\e^{-2\Psi} (1 + r\Psi^\prime) = m^\prime + 4\pi p r^2 = 4\pi r^2(\rho+p)
		\end{equation}
		When combined with the null energy condition, this takes the form, $r\Psi^\prime \ge -1$.},
	\begin{equation}\label{trapnullcond2}
		rp^\prime + p + \rho = \frac{qq^\prime}{4\pi r^3}  \quad {\rm or} \quad rp^\prime + p + \rho = \frac{\epsilon^\dprime}{8\pi r} + \frac{\epsilon^\prime}{4\pi r^2}
	\end{equation}
	
\begin{figure}[H]
        \centering
        \includegraphics[width=\linewidth]{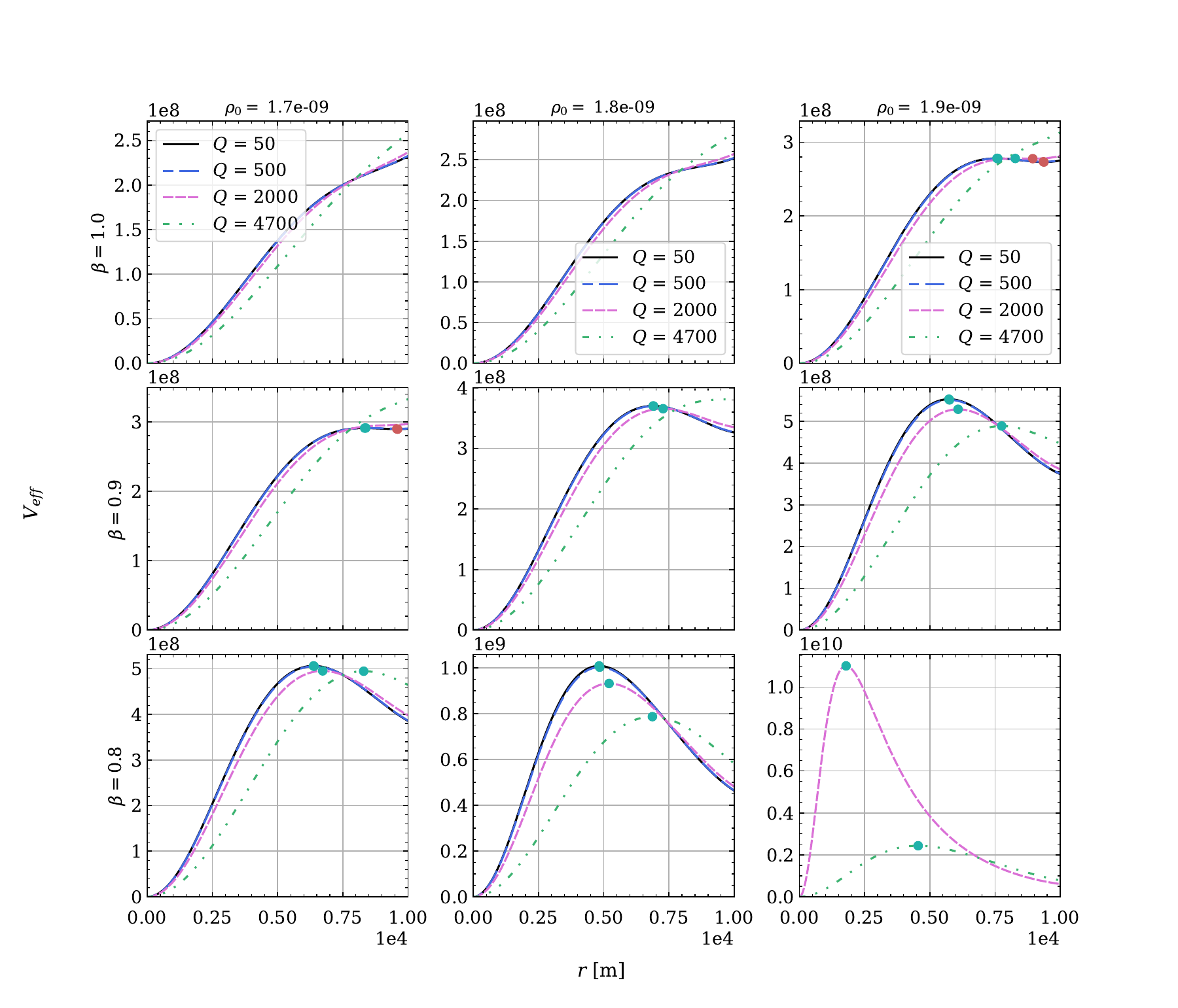}
        \caption{The effective potential, $V_{\rm eff}(r)$ plotted against the radial coordinate, $r$, for $K>0$. The green dots represent stable orbits while the red dots represent unstable orbits.}
        \label{fig-veff_kg0}
    \end{figure}

\begin{figure}[H]
        \centering
        \includegraphics[width=\linewidth]{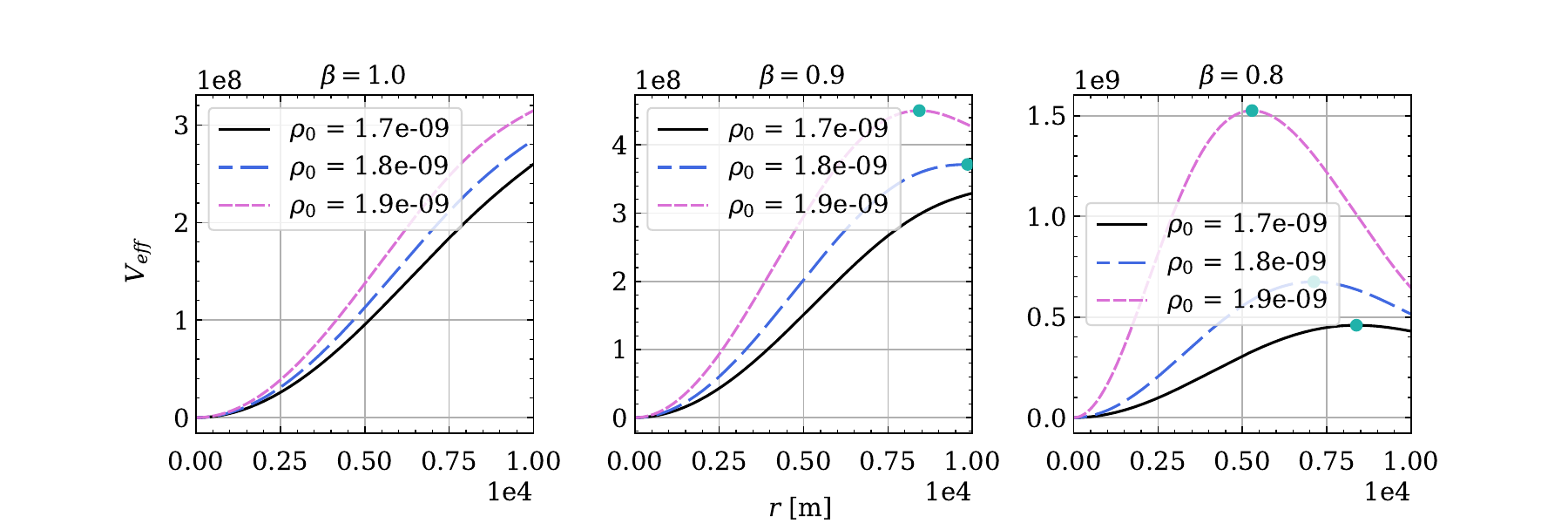}
        \caption{The effective potential, $V_{\rm eff}(r)$ plotted against the radial coordinate, $r$, for $K=0$. The green dots represent stable orbits while the red dots represent unstable orbits.}
        \label{fig-veff_ke0}
    \end{figure}

\begin{figure}[H]
        \centering
        \includegraphics[width=\linewidth]{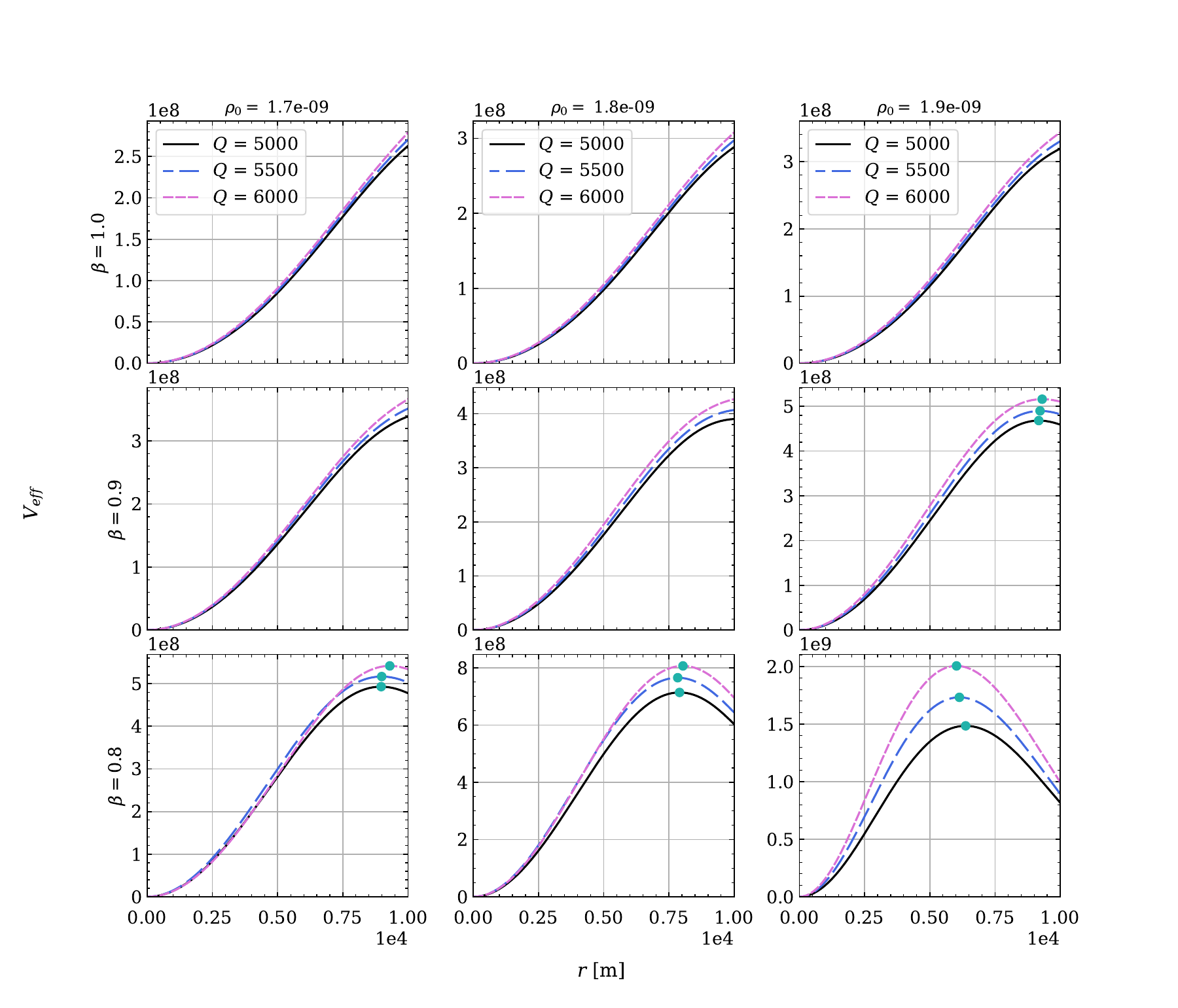}
        \caption{The effective potential, $V_{\rm eff}(r)$ plotted against the radial coordinate, $r$, for $K<0$. The green dots represent stable orbits while the red dots represent unstable orbits.}
        \label{fig-veff_kl0}
    \end{figure}

\begin{figure}[H]
        \centering
        \includegraphics[width=\linewidth]{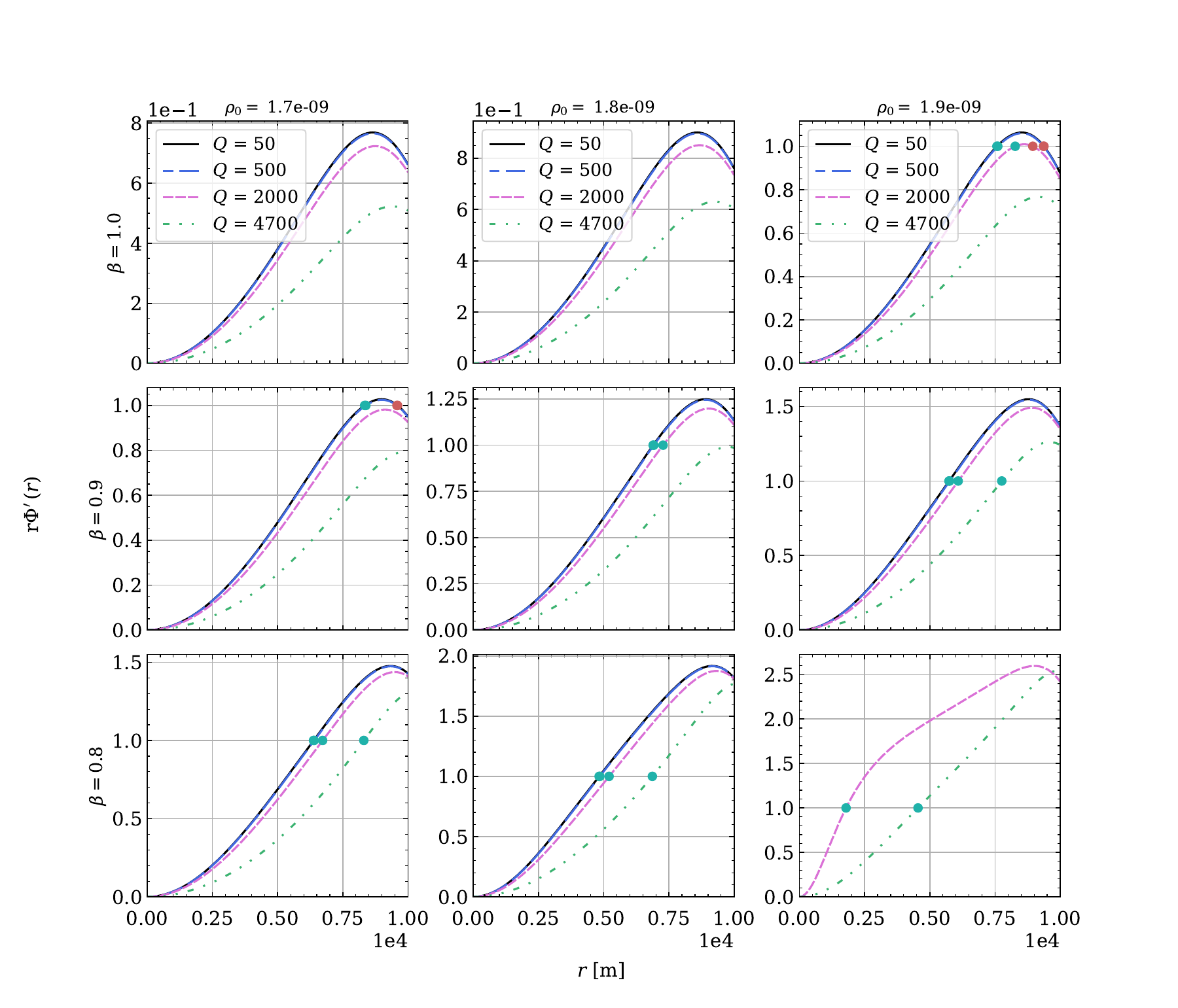}
        \caption{The function, $r\Phi^\prime$ plotted against the radial coordinate, $r$, for $K>0$. The green dots represent stable orbits while the red dots represent unstable orbits.}
        \label{fig-rphiprimekg0}
    \end{figure}

\begin{figure}[H]
        \centering
        \includegraphics[width=\linewidth]{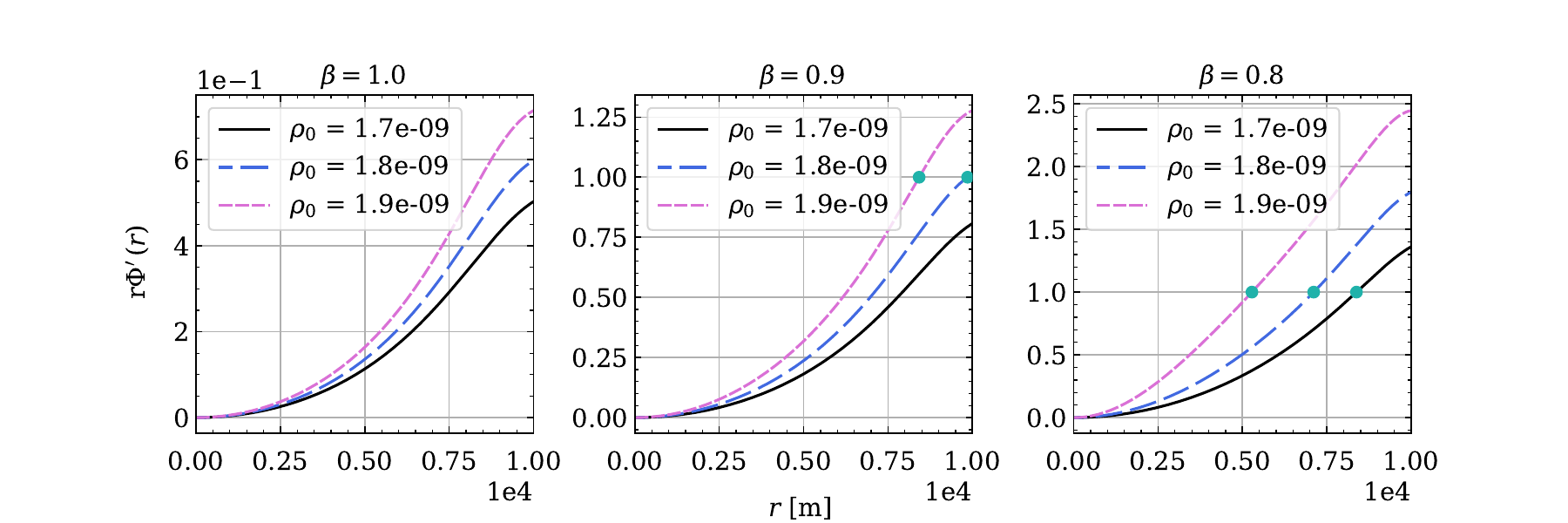}
        \caption{The function, $r\Phi^\prime$ plotted against the radial coordinate, $r$, for $K=0$. The green dots represent stable orbits while the red dots represent unstable orbits.}
        \label{fig-rphiprimeke0}
    \end{figure}

\begin{figure}[H]
            \centering
            \includegraphics[width=\linewidth]{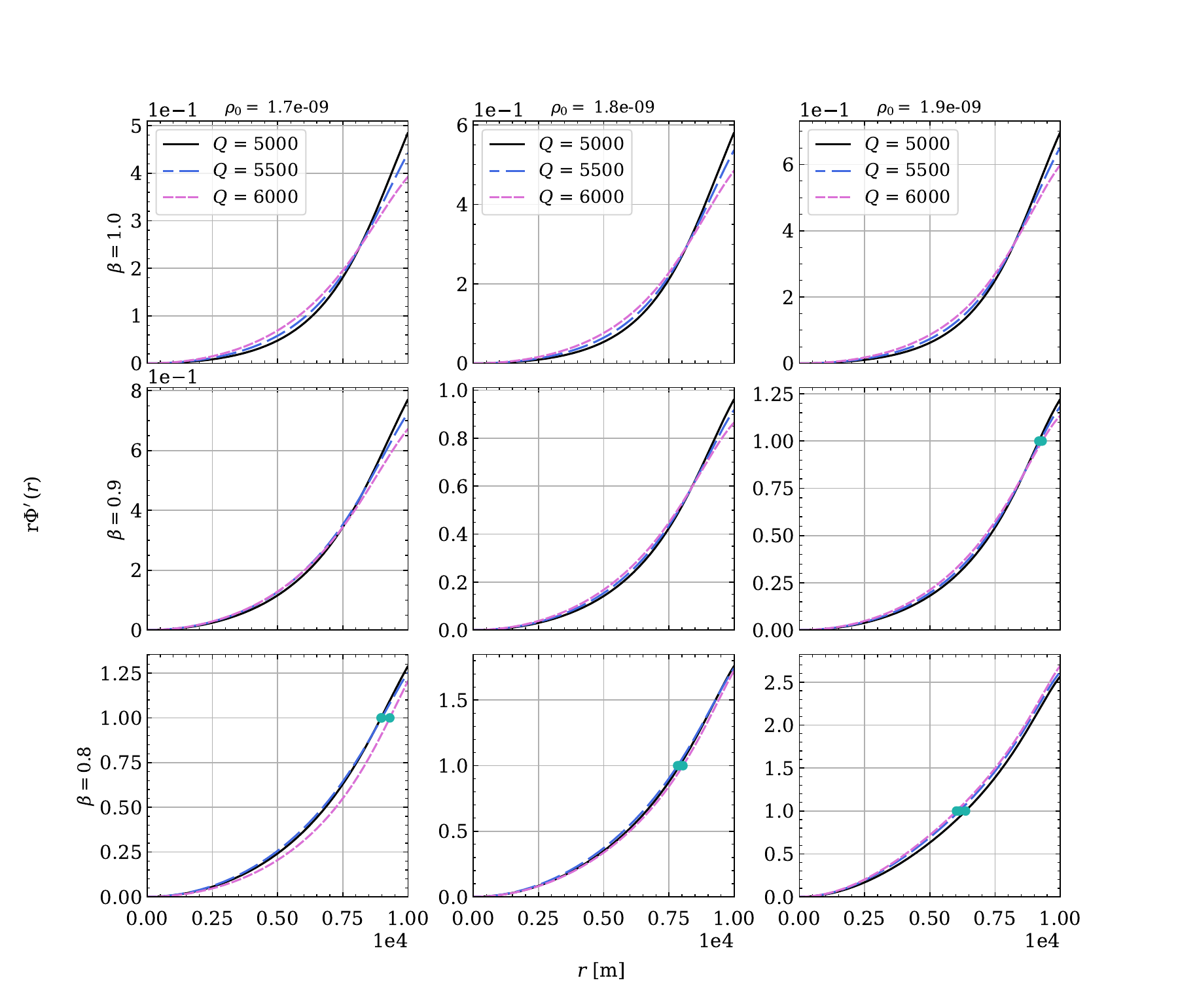}
            \caption{The function, $r\Phi^\prime$ plotted against the radial coordinate, $r$, for $K<0$. The green dots represent stable orbits while the red dots represent unstable orbits.}
            \label{fig-rphiprimekl0}
        \end{figure}

	Using equation \eqref{vefftol7}, the condition for existence of trapped null orbits in the generalised Tolman VII space-time becomes,
	\begin{equation}\label{trapnullcond3}
		\begin{rcases}
			&{\rm if}\ K>0 \quad \frac{r D w^\prime - \tan\left[D(w-w_b)\right]}{ 1 + rDw^\prime \tan\left[D(w - w_b)\right]}  \\
			&{\rm if}\ K = 0 \quad  D\left(r w^\prime - w + w_b\right)  \\
			&{\rm if}\ K<0 \quad \frac{r D w^\prime - \tanh\left[D(w-w_b)\right]}{ 1 - rDw^\prime \tanh\left[D(w - w_b)\right]} 
		\end{rcases}
		= \frac{\left(1 - \frac{Ar_b^2}{3} + \frac{Br_b^4}{5}\right)^{\hf}}{\frac{1}{D}\sqrt{\frac{5}{B}}\left\{\frac{B}{5}r_b^2 - \frac{A}{6} + 2\pi \rho_0(1-\beta) \right\}}
	\end{equation} 
	In figures \ref{fig-veff_kg0}, \ref{fig-veff_ke0}, \ref{fig-veff_kl0}, \ref{fig-rphiprimekg0}, \ref{fig-rphiprimeke0}, and \ref{fig-rphiprimekl0} we plot the effective potential and the function, $r\Phi^\prime$, for different values of the central density and $\beta$ for the three subclasses of the solution. The latter three figures match exactly with the former three thus establishing the equivalence between the conditions in equations \eqref{trapnullcond} and \eqref{trapnullcond0}.

\begin{figure}[H]
            \centering
    \includegraphics[width=\linewidth]{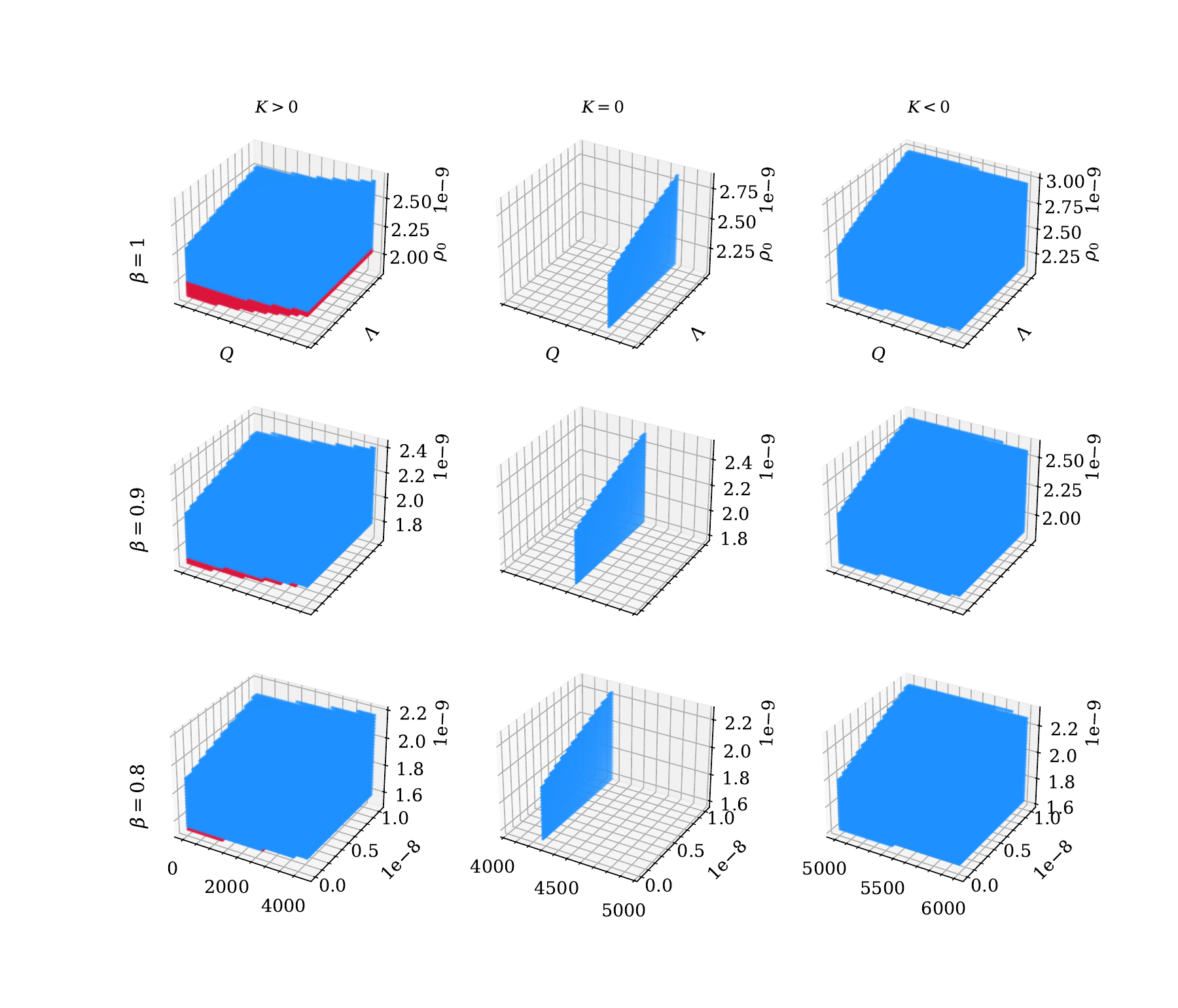}
            \caption{A plot exploring the three-dimensional parameter space of $\rho_0$, $Q$, and $\Lambda$. The region shaded blue corresponds to having only stable orbits while red corresponds both stable and unstable orbits. }
            \label{fig-Lambdavsrho0vsQ}
        \end{figure}

\begin{figure}[H]
            \centering
            \includegraphics[width=\linewidth]{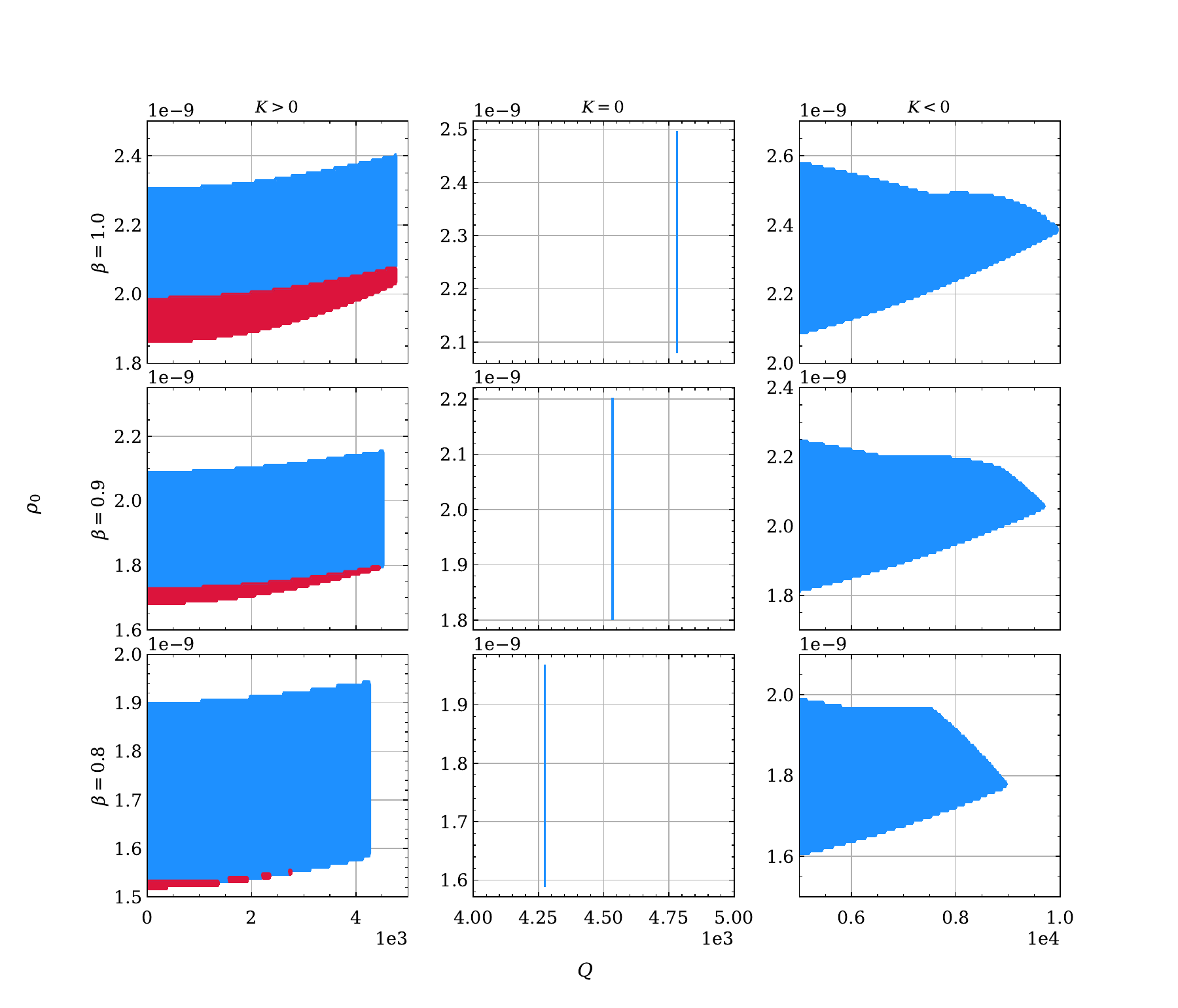}
            \caption{A plot exploring the two-dimensional parameter space of $\rho_0$ and $Q$. The region shaded blue corresponds to having only stable orbits while red corresponds both stable and unstable orbits. This figure represents a cross-sectional slice of figure \ref{fig-Lambdavsrho0vsQ} at $\Lambda = 10^{-52}$ m$^{-2}$.}
            \label{fig-Qvsrho_0}
        \end{figure}

\begin{figure}[H]
            \centering
            \includegraphics[width=\linewidth]{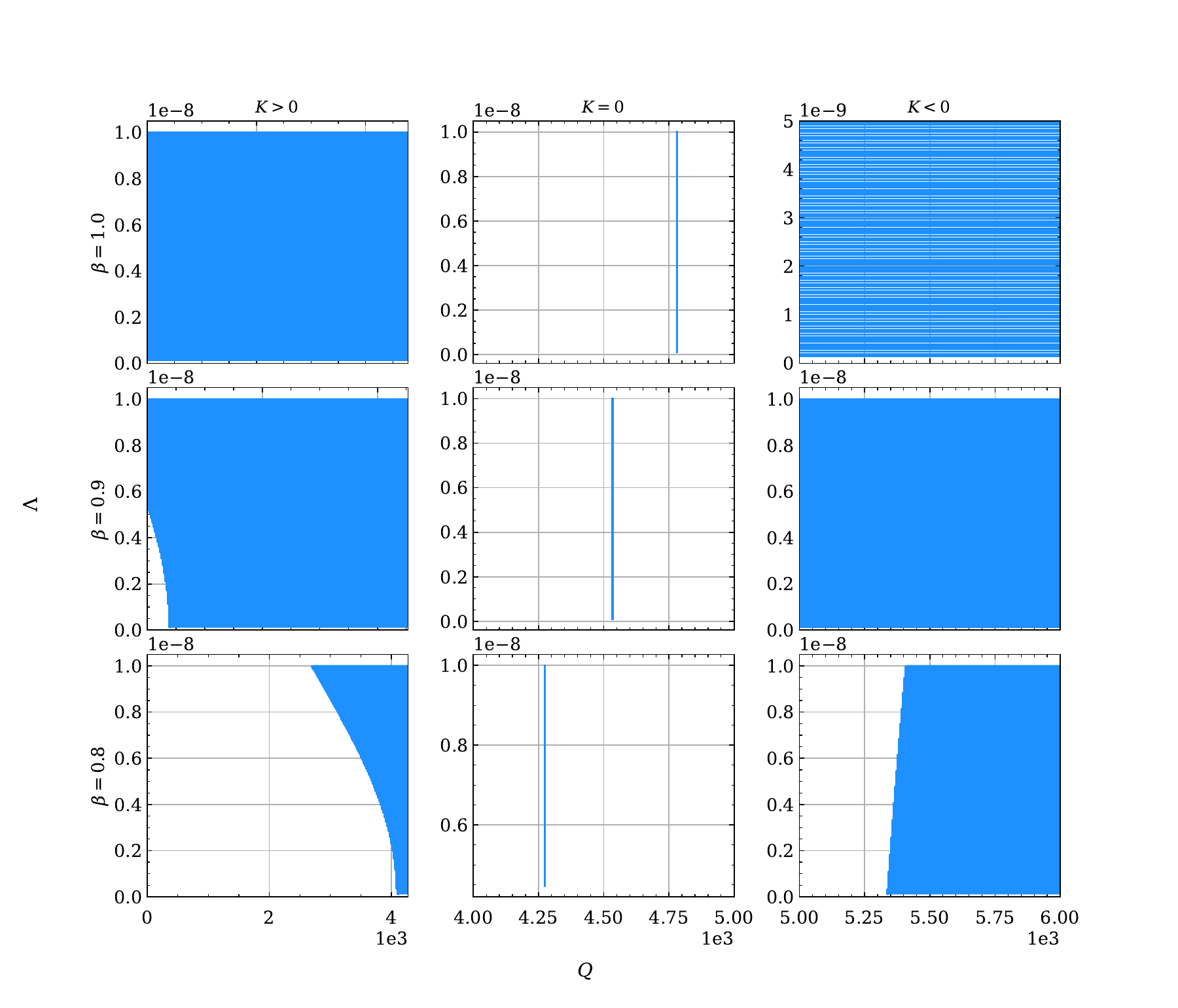}
            \caption{A plot exploring the two-dimensional parameter space of $\Lambda$ and $Q$. The region shaded blue corresponds to having only stable orbits while red corresponds both stable and unstable orbits. This figure represents a cross-sectional slice of figure \ref{fig-Lambdavsrho0vsQ} at $\rho_0 = 2\times 10^{-9}$ m$^{-2}$. }
            \label{fig-QvsLambda}
        \end{figure}

\begin{figure}[H]
            \centering
            \includegraphics[width=\linewidth]{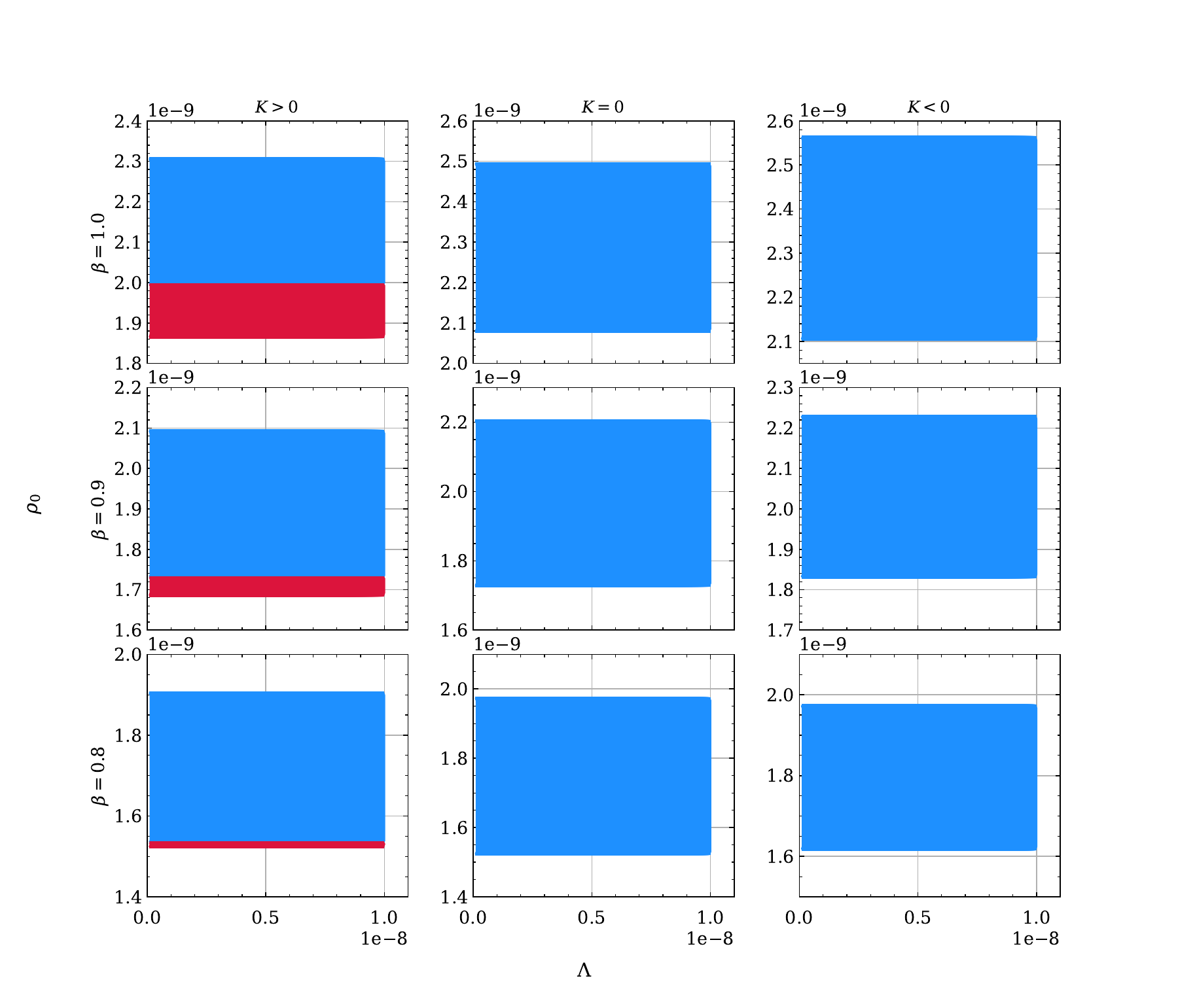}
            \caption{A plot exploring the two-dimensional parameter space of $\rho_0$ and $\Lambda$. The region shaded blue corresponds to having only stable orbits while red corresponds both stable and unstable orbits. This figure represents a cross-sectional slice of figure \ref{fig-Lambdavsrho0vsQ} at $Q = 1000$ m for $K>0$; $Q = 4780\ {\rm m},\ 4534\ {\rm m},\ 4275\ {\rm m}$ for $K = 0$ and $\beta = 1.0,\ 0.9,\ 0.8$, respectively; and $Q = 5500$ m for $K<0$.}
            \label{fig-Lambdavsrho0}
        \end{figure}

	These figures are made for representative values of various parameters to demonstrate that all three subclasses of the solution allow for trapping of null geodesics. In these figures, we note that for each subclass, given a value of $\beta$, trapped null geodesics exist for a range of values for $\rho_0, Q, \Lambda$. Therefore, to see the full extent of geodesic trapping, we need to explore the full parameter space. This is rather difficult due to the existence of five independent parameters. In the figures \ref{fig-Lambdavsrho0vsQ}, \ref{fig-Qvsrho_0}, \ref{fig-QvsLambda}, and \ref{fig-Lambdavsrho0} we explore the full parameter space. In each figure, the shaded area represents the region in the parameter space that allows for trapped null geodesics.

        From figure \ref{fig-Lambdavsrho0vsQ}, we see that a combination all five free parameters of the system determine the trapping of null geodesics. Through exploring this parameter space in its entirety, we find that the Tolman VII solution allows for trapped null orbits for tenuity $\left(\frac{r_b}{m(r_b)}\right)$ values between approximately, 1.93 and 3.06 for the total charge, $0<Q<6000$.
        
        Further, realistic values of the cosmological constant ($\sim 10^{-52}$ m$^{-2}$) do not affect the trapping. Only when the  value of the cosmological constant is comparable to the central density, $\rho_0$, does it modify the trapping of null geodesics. This makes sense since it is really the constant, $A = 8\pi \rho_0 - \Lambda$ that determines whether trapped orbits will exist or not. In the figures that follow, we plot three different cross-sections of the above three-dimensional plot as an exposition into the parameter space relevant for trapping.

	\section{Physical Properties of the Fluid} \label{sec-physprop}
	\subsection{Radial Profile of the Pressure}
\begin{figure}[H]
            \centering
            \includegraphics[width=\linewidth]{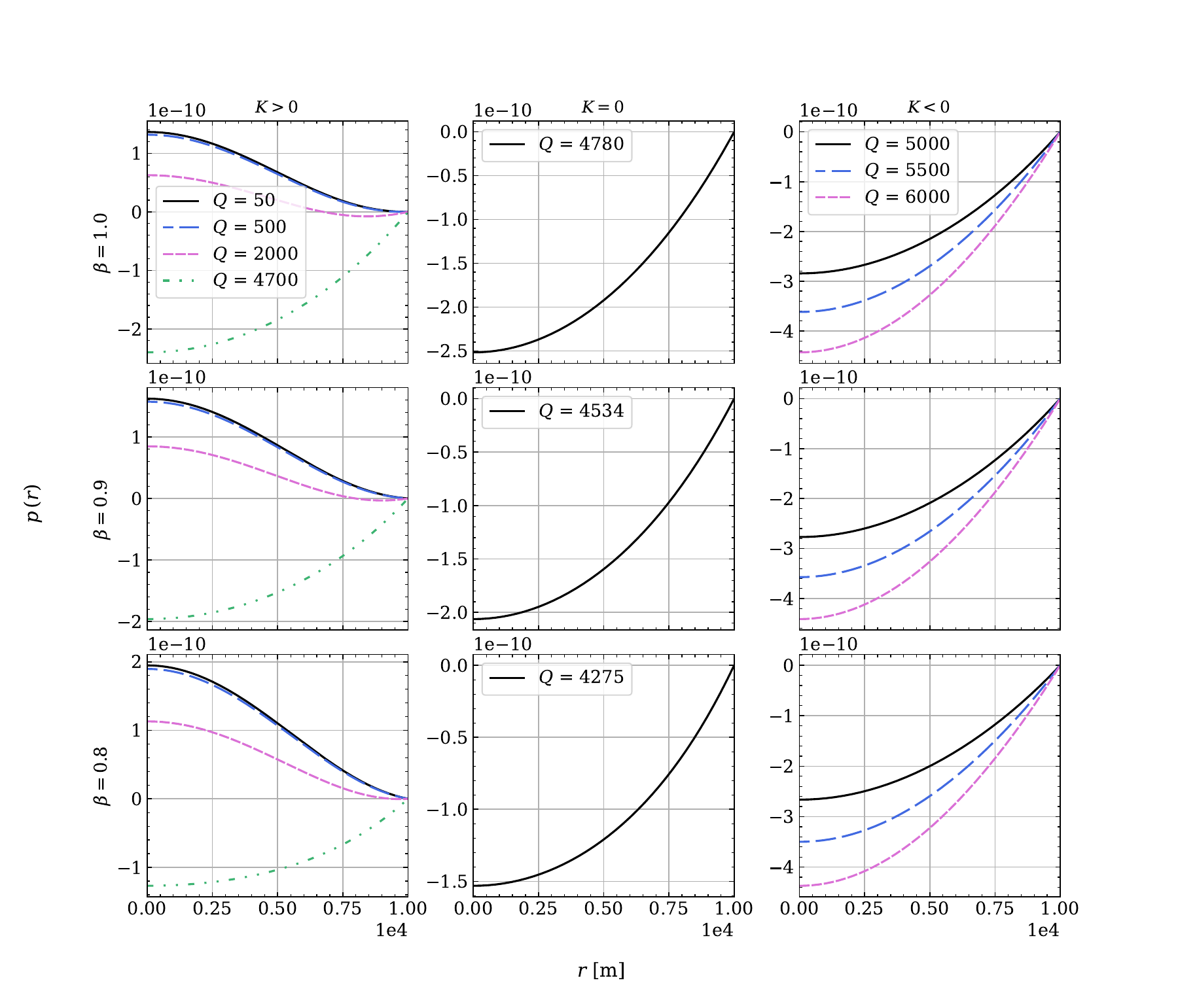}
            \caption{The pressure, $p(r)$, plotted as a function of the radial coordinate, $r$. The values of various parameters are listed in table \ref{tab-paramval}. The pressure is only monotonically decreasing for smaller values of $Q$ in the case, $K>0$. This makes this case physically more relevant than other cases.}
            \label{fig-press}
        \end{figure}

	Using equation \eqref{tolconst} in \eqref{press3}, we get the full expression for the radial profile of the pressure,
		\begin{equation}\label{press4}
			p(r) = 
			\begin{cases}
				\scalemath{0.6}{\frac{D}{2\pi}\sqrt{\frac{B}{5}} \left(1 - \frac{Ar^2}{3} + \frac{Br^4}{5}\right)^\hf \left[ \frac{1 - \frac{\left(1 - \frac{Ar_b^2}{3} + \frac{Br_b^4}{5}\right)^{\hf}}{\frac{1}{D}\sqrt{\frac{5}{B}}\left\{\frac{B}{5}r_b^2 - \frac{A}{6} + 2\pi\rho_0(1-\beta)\right\}} \tan[D(w-w_b)]}{\frac{\left(1 - \frac{Ar_b^2}{3} + \frac{Br_b^4}{5}\right)^{\hf}}{\frac{1}{D}\sqrt{\frac{5}{B}}\left\{\frac{B}{5}r_b^2 - \frac{A}{6} + 2\pi\rho_0(1-\beta)\right\}} + \tan[D(w-w_b)]} \right] - \frac{1}{2\pi} \left(\frac{B}{5}r^2 - \frac{A}{6}\right) - \rho_0 \left[ 1 - \beta \left( \frac{r}{r_b} \right)^2\right] }  \quad &{\rm if}\ K>0 \\ 
				\scalemath{0.58}{\frac{D}{2\pi}\sqrt{\frac{B}{5}} \left(1 - \frac{Ar^2}{3} + \frac{Br^4}{5}\right)^\hf \left[ \frac{\left(1 - \frac{Ar_b^2}{3} + \frac{Br_b^4}{5}\right)^{\hf}}{\frac{1}{D}\sqrt{\frac{5}{B}}\left\{\frac{B}{5}r_b^2 - \frac{A}{6} + 2\pi\rho_0(1-\beta)\right\}} + D(w-w_b) \right]^{-1} - \frac{1}{2\pi} \left(\frac{B}{5}r^2 - \frac{A}{6}\right) - \rho_0 \left[ 1 - \beta \left( \frac{r}{r_b} \right)^2\right]} \quad &{\rm if}\ K=0\\
				\scalemath{0.6}{\frac{D}{2\pi}\sqrt{\frac{B}{5}} \left(1 - \frac{Ar^2}{3} + \frac{Br^4}{5}\right)^\hf \left[ \frac{1 + \frac{\left(1 - \frac{Ar_b^2}{3} + \frac{Br_b^4}{5}\right)^{\hf}}{\frac{1}{D}\sqrt{\frac{5}{B}}\left\{\frac{B}{5}r_b^2 - \frac{A}{6} + 2\pi\rho_0(1-\beta)\right\}} \tanh[D(w-w_b)]}{\frac{\left(1 - \frac{Ar_b^2}{3} + \frac{Br_b^4}{5}\right)^{\hf}}{\frac{1}{D}\sqrt{\frac{5}{B}}\left\{\frac{B}{5}r_b^2 - \frac{A}{6} + 2\pi\rho_0(1-\beta)\right\}} + \tanh[D(w-w_b)]} \right] - \frac{1}{2\pi} \left(\frac{B}{5}r^2 - \frac{A}{6}\right) - \rho_0 \left[ 1 - \beta \left( \frac{r}{r_b} \right)^2\right] } \quad &{\rm if}\ K<0
			\end{cases}
		\end{equation}
	Having derived its analytical expression, we plot the pressure as a function of the radius in figure \ref{fig-press}.
        We note that the pressure is neither monotonically decreasing nor does it remain positive everywhere for the subclasses corresponding to $K \le 0$. Therefore, from figure \ref{fig-press}, we can conclude that the case with $K>0$ is the only physically relevant subclass of the solution. Moreover, even in this subclass, the monotonically decreasing nature of pressure is only present for smaller values of the total charge, $Q$.

    \subsection{Equation of State}
    In order to gauge the physical properties of the fluid even more, it is useful to find an equation of state, i.e., the pressure as a function of the density. In the case of Tolman VII solution, this is rather straightforward -- one can simply invert equation \eqref{rhotol} and replace, $r^2 = \frac{r_b^2}{\rho_0\beta}\left(\rho_0 - \rho\right)$ in equation \eqref{press4}. We do this, and after performing some rather cumbersome algebra, we find the equation of state to be,
	\begin{equation}\label{eos}
		p(\rho) = 
		\begin{cases}
				\scalemath{0.5}{\frac{D}{2\pi}\sqrt{\frac{B}{5}} \left\{1 - \frac{Ar_b^2}{3\rho_0\beta} \left(\rho_0 - \rho\right) + \frac{Br_b^4}{5\beta^2\rho_0^2}(\rho_0 - \rho)^2\right\}^\hf \left[ \frac{1 - \frac{\left(1 - \frac{Ar_b^2}{3} + \frac{Br_b^4}{5}\right)^{\hf}}{\frac{1}{D}\sqrt{\frac{5}{B}}\left\{\frac{B}{5}r_b^2 - \frac{A}{6} + 2\pi\rho_0(1-\beta)\right\}} \tan[D(w-w_b)]}{\frac{\left(1 - \frac{Ar_b^2}{3} + \frac{Br_b^4}{5}\right)^{\hf}}{\frac{1}{D}\sqrt{\frac{5}{B}}\left\{\frac{B}{5}r_b^2 - \frac{A}{6} + 2\pi\rho_0(1-\beta)\right\}} + \tan[D(w-w_b)]} \right] - \frac{1}{2\pi} \left(\frac{Br_b^2}{5\rho_0\beta}(\rho_0 - \rho) - \frac{A}{6}\right) - \rho} \quad &{\rm if}\ K>0\\
				\scalemath{0.5}{\frac{D}{2\pi}\sqrt{\frac{B}{5}} \left\{1 - \frac{Ar_b^2}{3\rho_0\beta} \left(\rho_0 - \rho\right) + \frac{Br_b^4}{5\beta^2\rho_0^2}(\rho_0 - \rho)^2\right\}^\hf \left[ \frac{\left(1 - \frac{Ar_b^2}{3} + \frac{Br_b^4}{5}\right)^{\hf}}{\frac{1}{D}\sqrt{\frac{5}{B}}\left\{\frac{B}{5}r_b^2 - \frac{A}{6} + 2\pi\rho_0(1-\beta)\right\}} + D(w-w_b) \right]^{-1} - \frac{1}{2\pi} \left(\frac{Br_b^2}{5\rho_0\beta}(\rho_0 - \rho) - \frac{A}{6}\right) - \rho}  \quad &{\rm if}\ K=0\\
				\scalemath{0.5}{\frac{D}{2\pi}\sqrt{\frac{B}{5}} \left\{1 - \frac{Ar_b^2}{3\rho_0\beta} \left(\rho_0 - \rho\right) + \frac{Br_b^4}{5\beta^2\rho_0^2}(\rho_0 - \rho)^2\right\}^\hf \left[ \frac{1 + \frac{\left(1 - \frac{Ar_b^2}{3} + \frac{Br_b^4}{5}\right)^{\hf}}{\frac{1}{D}\sqrt{\frac{5}{B}}\left\{\frac{B}{5}r_b^2 - \frac{A}{6} + 2\pi\rho_0(1-\beta)\right\}} \tanh[D(w-w_b)]}{\frac{\left(1 - \frac{Ar_b^2}{3} + \frac{Br_b^4}{5}\right)^{\hf}}{\frac{1}{D}\sqrt{\frac{5}{B}}\left\{\frac{B}{5}r_b^2 - \frac{A}{6} + 2\pi\rho_0(1-\beta)\right\}} + \tanh[D(w-w_b)]} \right] - \frac{1}{2\pi} \left(\frac{Br_b^2}{5\rho_0\beta}(\rho_0 - \rho) - \frac{A}{6}\right) - \rho } \quad &{\rm if}\ K<0
		\end{cases}
	\end{equation}
	where, we have,
	\begin{equation}
		w(\rho) = \ln\left[ \left\{ \frac{15\rho_0\beta - 5Ar_b^2(\rho_0 - \rho)}{3B\rho_0\beta} + \frac{r_b^4}{\rho_0^2\beta^2}(\rho_0 - \rho)^2 \right\}^\hf + \frac{r_b^2}{\rho_0\beta}(\rho_0 - \rho) - \frac{5A}{6B} \right]
	\end{equation}
	

    \begin{figure}[H]
        \centering
        \includegraphics[width=\linewidth]{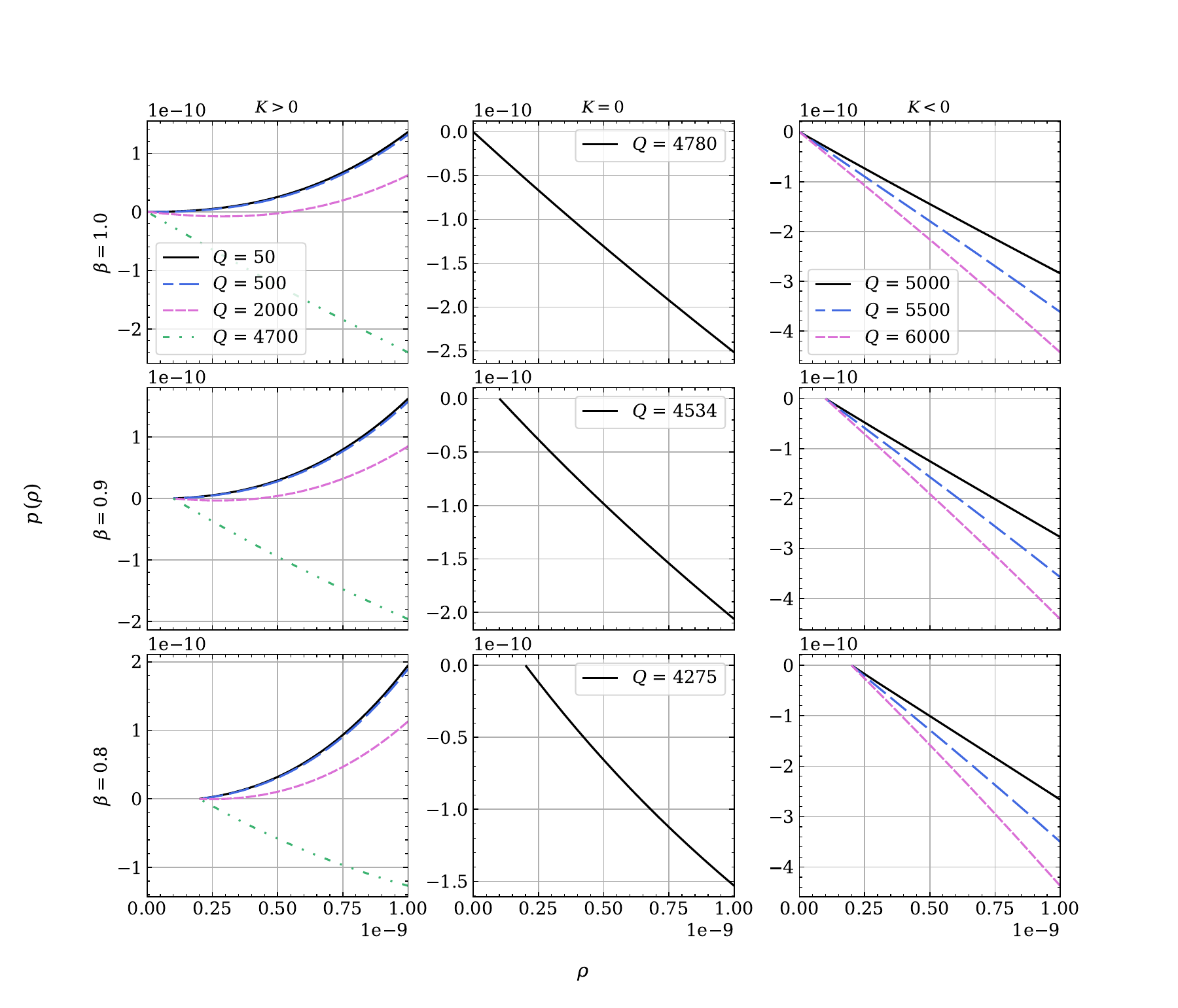}
        \caption{The pressure, $p(\rho)$, plotted as a function of the density, $\rho$. The values of various parameters are listed in table \ref{tab-paramval}. As expected, the equation of state is well-behaving only for smaller values of $Q$ in the case, $K>0$.}
        \label{fig-eos}
    \end{figure}

    \subsubsection*{\normalsize Polytropic Index}
\begin{figure}[H]
        \centering
        \includegraphics[width=\linewidth]{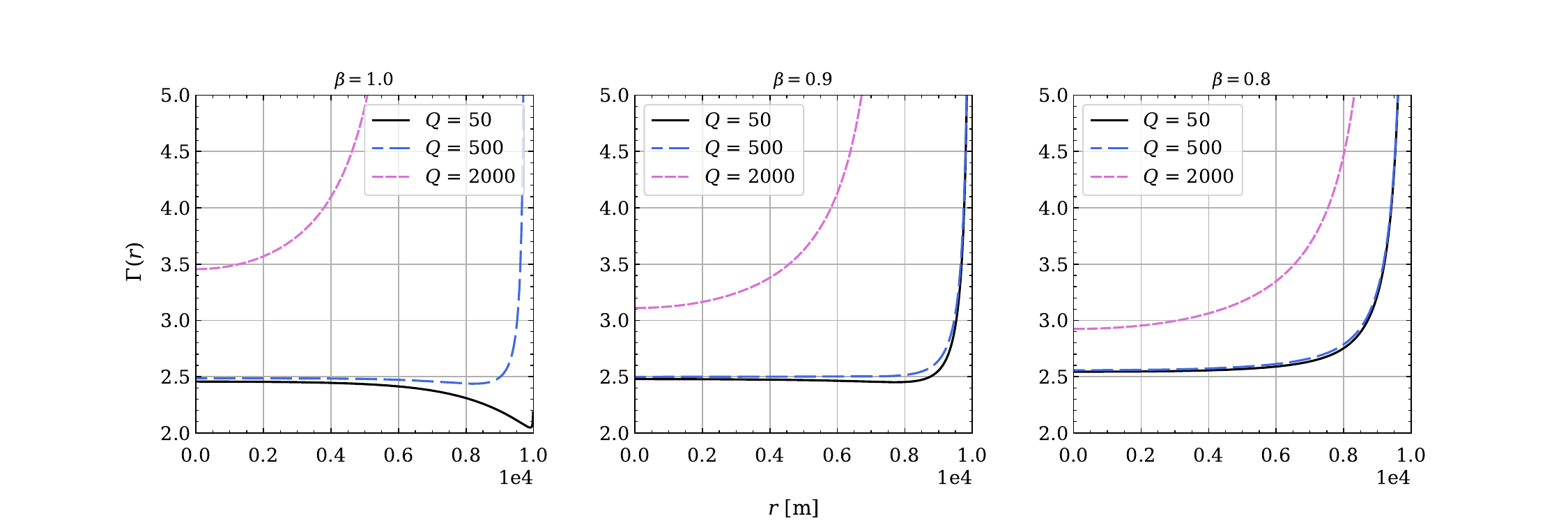}
        \caption{The polytropic index, $\Gamma(r)= \frac{\partial \log p}{\partial \log \rho}$, plotted as a function of the radial coordinate, $r$. The values of various parameters are listed in table \ref{tab-paramval}. For smaller values of $Q$, the Tolman VII fluid in the case, $K>0$, can be considered a polytrope with $\Gamma \sim 2.5$. The polytropic index diverges for high values of $Q$ even in the case with $K>0$. However, for $\beta<1$, a few high values of $Q$ are consistent with the polytropic index being, $\Gamma \sim 3.0$ if the boundary radius is taken to be, $r_b \sim 6000$ m.}
        \label{fig-polyind}
    \end{figure}

	Even though we are able to obtain an analytic expression for the equation of state, it is quite abstruse. Therefore, it is a useful exercise to fit the equation of state to a simpler form. We choose to do so assuming the fluid to be a polytrope, $p(\rho) = \kappa \rho^{\Gamma(r)}$. In figure \ref{fig-polyind}, we plot the polytropic index, $\Gamma(r) = \frac{\partial \log p}{\partial \log \rho}$, for the Tolman VII fluid in the case, $K>0$ (since this is the only physically relevant case). We find that the fluid can be considered to be a polytrope with an index, $\Gamma \sim 2.5$. This is consistent with previous such studies in the literature \cite{raghoo}. Additionally, we also note that higher values of the polytropic index, $\Gamma \sim 3.0$ could fit for higher values of $Q$, $\beta<1$ and $r_b \sim 6000$ m.

%
%
%
%
%
	
    \subsection{Sound Speed}
\begin{figure}[H]
            \centering
            \includegraphics[width=\linewidth]{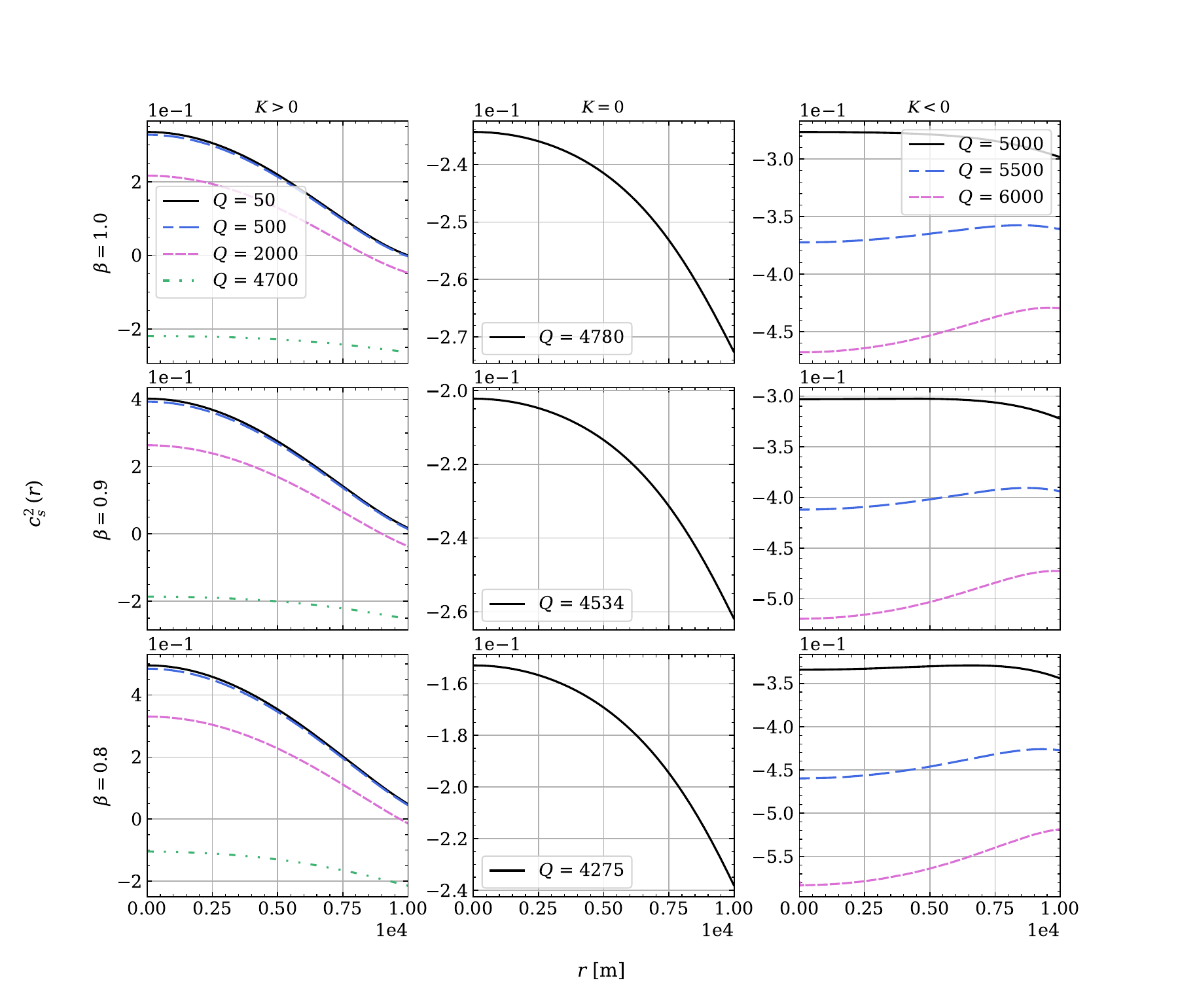}
            \caption{The sound speed, $c_s^2 = \frac{\partial p}{\partial\rho} $, plotted against the radial coordinate, $r$. The values of various parameter are given in table \ref{tab-paramval}. The subclasses $K\le 0$ have a negative sound speed thus making them physically irrelevant. For the subclass, $K>0$, on one hand, causality is violated for $\beta<0$ for small values of $Q$ while on the other hand, the sound speed becomes negative for $0.8<\beta<1.0$ and large values of $Q$. This restricts the self-boundedness parameter, $\beta$, to be between 0.8 and 1.0. }
            \label{fig-sspeed}
        \end{figure}

	Once, we have the equation of state, it is straightforward to calculate the sound speed, $c_s^2 = \frac{\partial p}{\partial\rho}$. The sound speed is plotted against the radial coordinate in figure \ref{fig-sspeed}. Once again, for the subclasses with $K\le 0$, the sound speed is negative. Moreover, we see that even for $K>0$, the sound speed either becomes greater than unity, thus violating causality (for small $Q$ and $\beta<0.8$) or becomes negative (for large $Q$ and $\beta <0.8$). Therefore, in addition to concluding that the cases with $K\le 0$ are physically irrelevant, the sound speed also puts a constraint on the values that the self-boundedness parameter, $\beta$, can take. This is the reason we do not present any analysis for $\beta<0.8$ throughout this paper. 
        

	\subsection{Energy Conditions}
\begin{figure}[H]
            \centering
            \includegraphics[width=\linewidth]{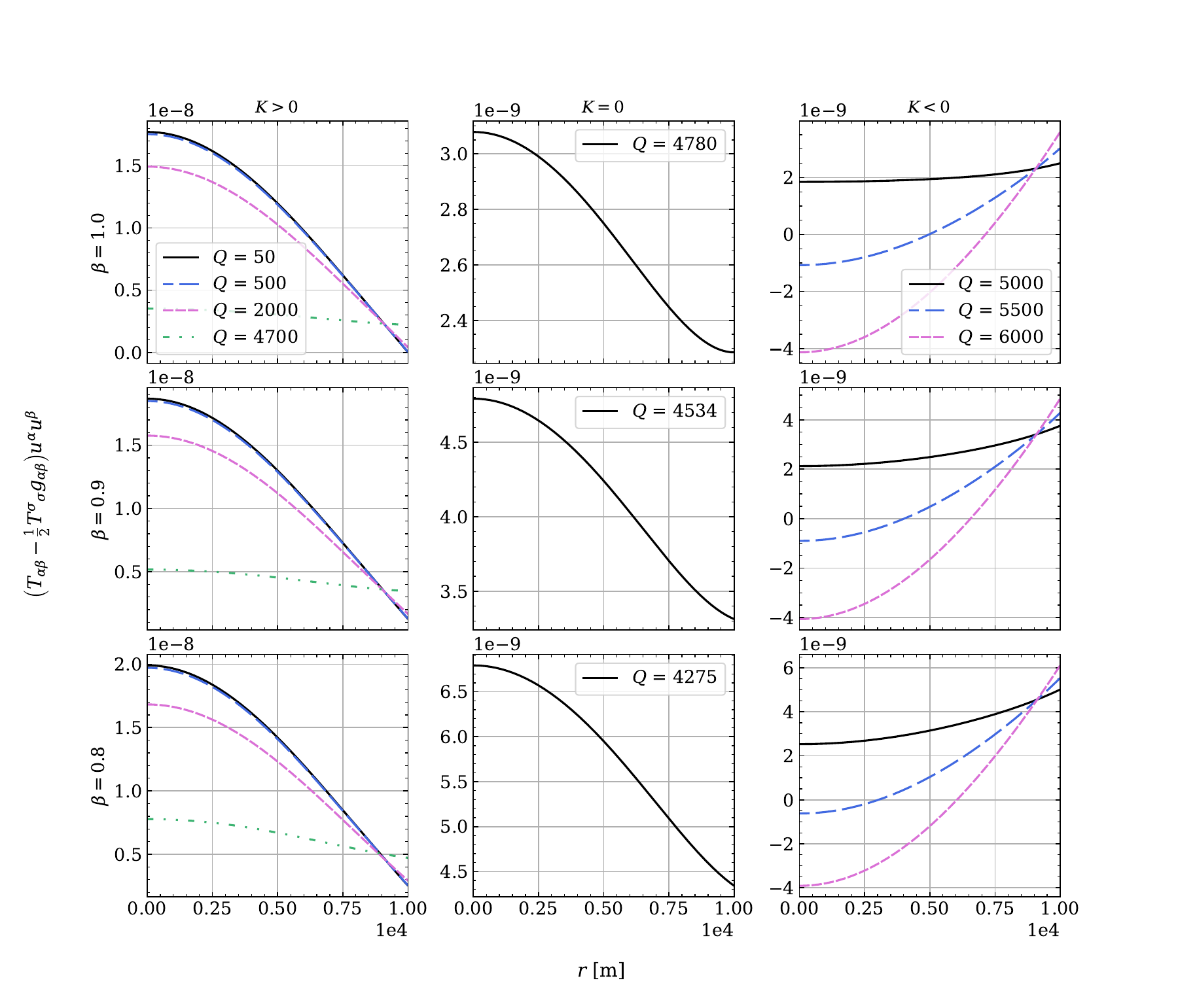}
            \caption{The quantity, $\left(T_{\alpha\beta} - \hf {T^\sigma}_\sigma g_{\alpha\beta}\right)u^\alpha u^\beta$, plotted against the radial coordinate, $r$. The values of various parameters is given in table \ref{tab-paramval}. For $K\ge 0$, this quantity remains non-negative at all values of $r$, thus showing that the strong energy condition is obeyed by these two subclasses of the solution. However, for the subclass with $K<0$, the cases with $Q> \sim 5500$ m violate the strong energy condition.}
            \label{fig-strongec}
        \end{figure}

	The energy conditions are constraints on the total energy-momentum tensor which, in our case, includes matter fields, electromagnetic fields, and the cosmological constant. Therefore, we have,
	\begin{equation}
		T_{\alpha\beta} =\ ^{(\rm matter)}T_{\alpha\beta}\ +\ ^{(\rm emfield)}T_{\alpha\beta} + \Lambda g_{\alpha\beta}
	\end{equation}
	\begin{enumerate}[itemsep=1mm,leftmargin=*]
		\item The null energy condition, $T_{\alpha\beta}k^\alpha k^\beta \ge 0$, reduces to,
		\begin{equation}
			\rho + p \ge 0
		\end{equation}
		where, $k^\alpha = \left[\e^{-\Phi}, \e^{-\Psi}, 0, 0\right]$ is a null vector. We see that the presence of a charge and a cosmological constant does not change the form of the null energy condition. Figure \ref{fig-nullec}, shows that all three subclasses of the solution obey the null energy condition.
        \begin{figure}[H]
            \centering
            \includegraphics[width=\linewidth]{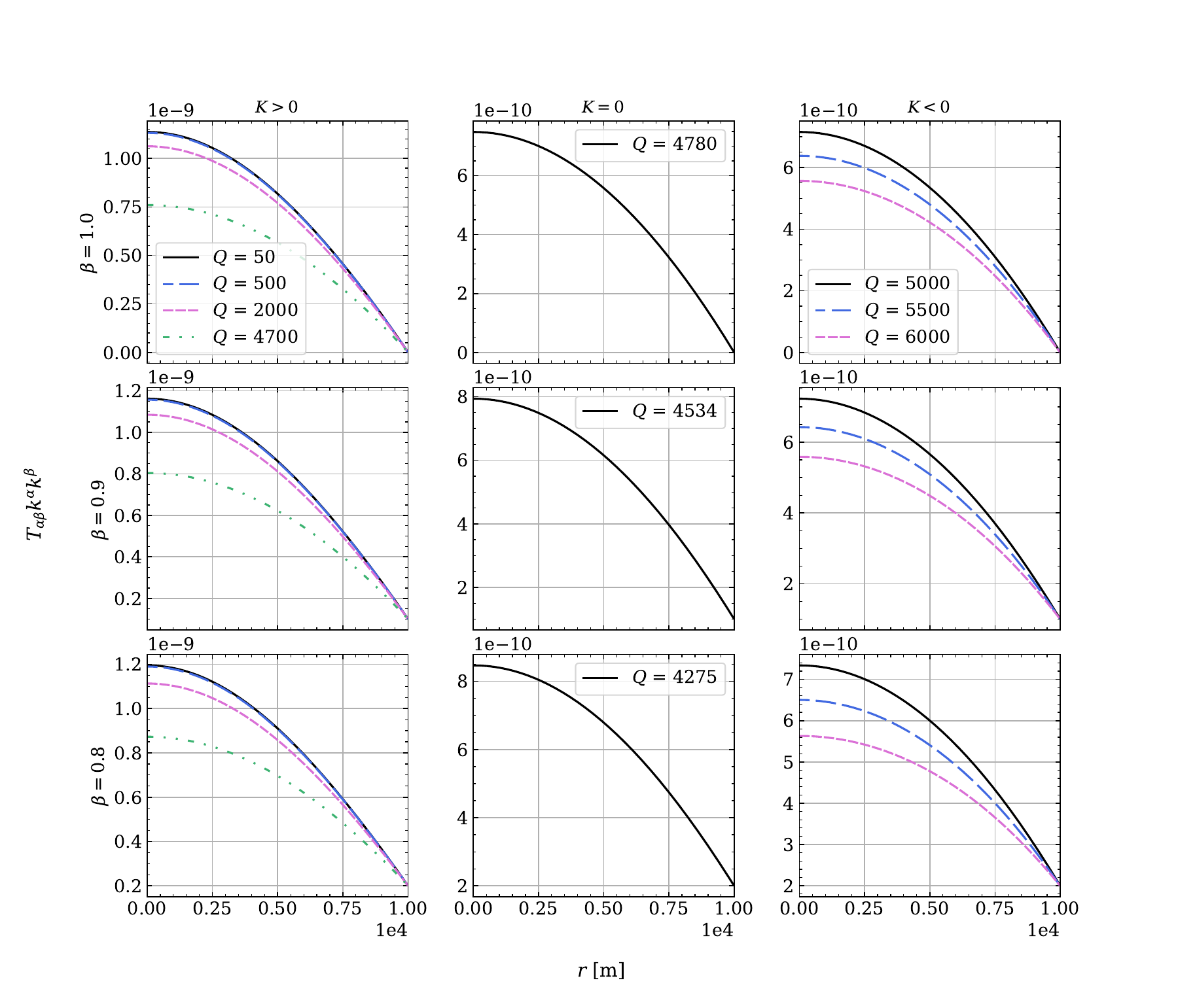}
            \caption{The quantity, $T_{\alpha\beta}k^\alpha k^\beta$, plotted against the radial coordinate, $r$. The values of various parameters is given in table \ref{tab-paramval}. This quantity remains non-negative at all values of $r$, thus showing that the null energy condition is obeyed by all three subclasses of the solution.}
            \label{fig-nullec}
        \end{figure}
		
		\item The weak energy condition, $T_{\alpha\beta}u^\alpha u^\beta \ge 0$, reduces to,
		\begin{equation}
			8\pi \rho + \frac{q^2}{r^4} - \Lambda \ge  0
		\end{equation}
		where, $u^\alpha = \left[\e^{-\Phi},0,0,0\right]$ is a timelike vector. Using the ansatzes \eqref{rhotol} and \eqref{chargetol}, the weak energy condition reduces to,
		\begin{equation}
			A - Br^2 \ge 0
		\end{equation}
		Comparing this with equation \eqref{poscurvcond1}, we can conclude that the weak energy condition ensures positivity of the spatial curvature in the generalised Tolman VII solution\footnote{In general, for any static spherically symmetric perfect fluid solution, the weak energy condition only ensures the non-negativity of the spatial curvature.}. Figure \ref{fig-weakec} shows that all three subclasses of the solution obey the weak energy condition.

        \begin{figure}[H]
            \centering
            \includegraphics[width=\linewidth]{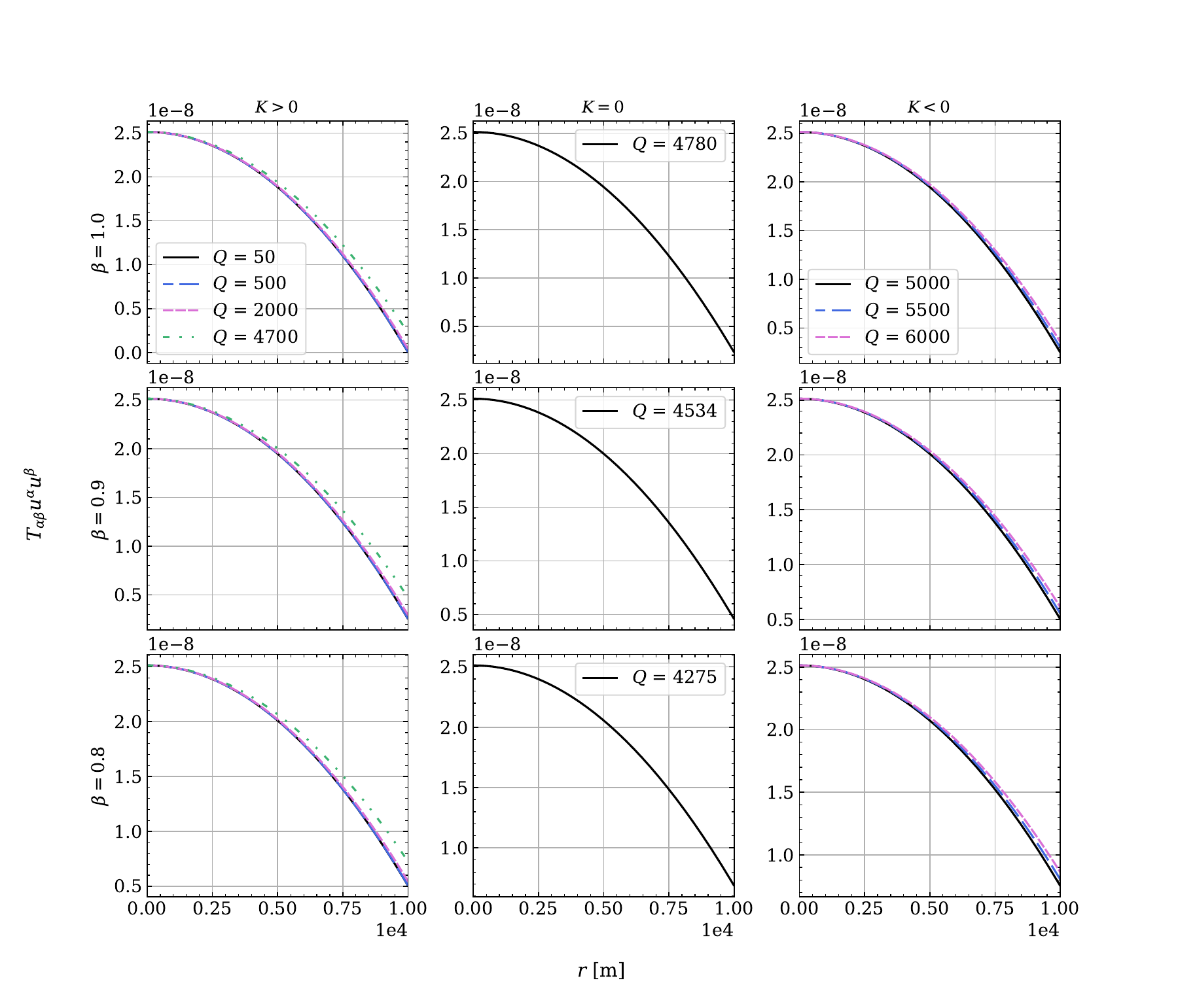}
            \caption{The quantity, $T_{\alpha\beta}u^\alpha u^\beta$, plotted against the radial coordinate, $r$. The values of various parameters is given in table \ref{tab-paramval}. This quantity remains non-negative at all values of $r$, thus showing that the weak energy condition is obeyed by all three subclasses of the solution.}
            \label{fig-weakec}
        \end{figure}
		
		\item The dominant energy condition, $T_{\alpha\beta}v^\beta {T^\alpha}_\sigma v^\sigma \le 0$, reduces to,
		\begin{align}
			8\pi \rho + \frac{q^2}{r^4} - \Lambda &\ge 0 \quad {\rm when}\ v^\alpha\ {\rm is\ timelike} \\
			4\pi (\rho - p) + \frac{q^2}{r^4} - \Lambda &\ge 0 \quad {\rm when}\ v^\alpha\ {\rm is\ null}
		\end{align}
		The timelike case is the same as the weak energy condition, therefore, the null case is the only independent condition. Figure \ref{fig-domec} (along with figure \ref{fig-weakec}) shows that all three subclasses of the solution obey the dominant energy condition.

        \begin{figure}[H]
            \centering
            \includegraphics[width=\linewidth]{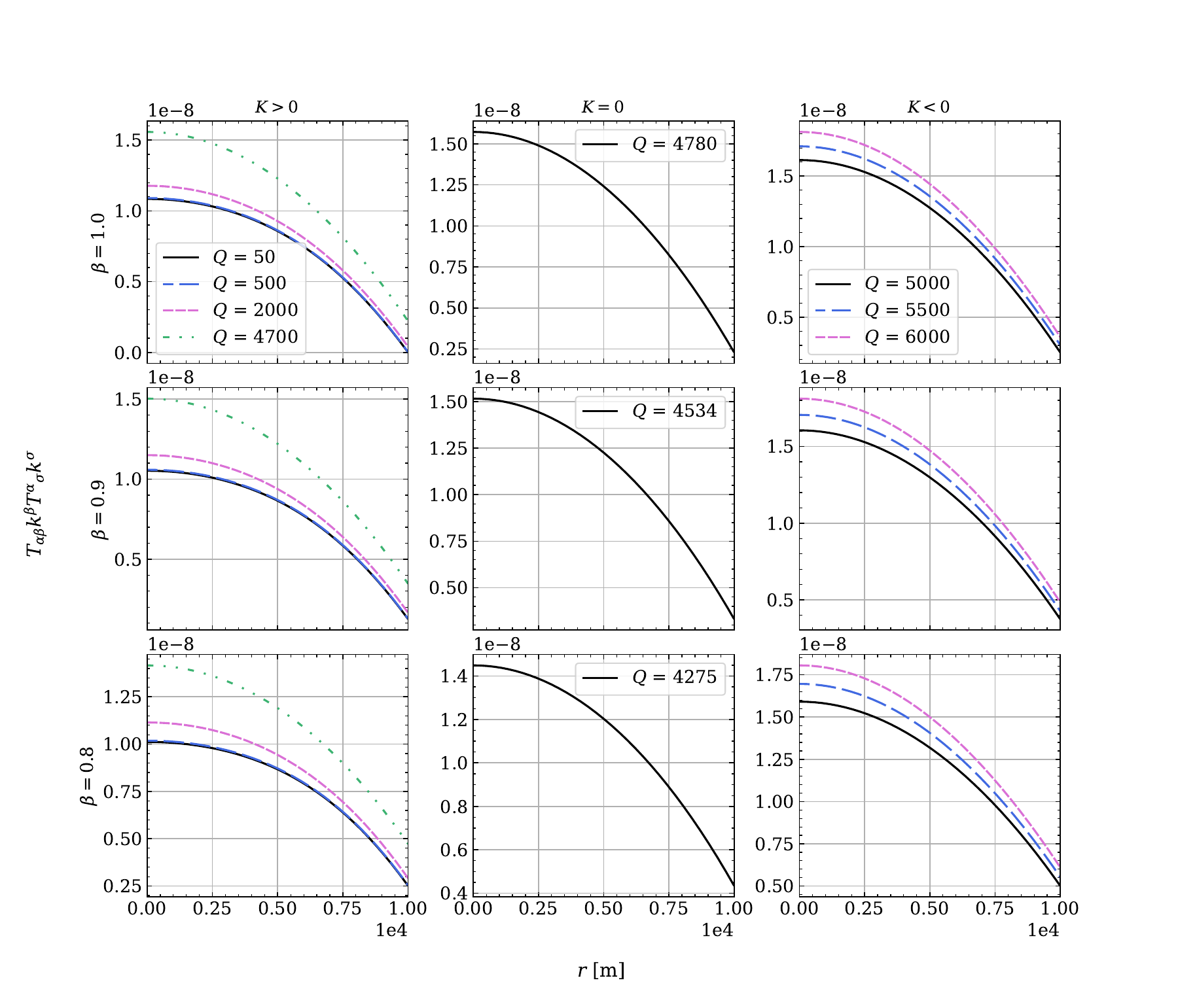}
            \caption{The quantity, $T_{\alpha\beta}v^\beta {T^\alpha}_\sigma v^\sigma$, plotted against the radial coordinate, $r$,, assuming the vector, $v^\alpha$, is null. The values of various parameters is given in table \ref{tab-paramval}. This quantity remains non-negative at all values of $r$, thus showing that the dominant energy condition is obeyed by all three subclasses of the solution.}
            \label{fig-domec}
        \end{figure}
		
		\item The strong energy condition, $\left(T_{\alpha\beta} - \hf {T^\sigma}_\sigma g_{\alpha\beta}\right)u^\alpha u^\beta \ge 0$, reduces to,
		\begin{equation}
			4\pi (\rho + 3p) + \frac{q^2}{r^4} + \Lambda \ge 0
		\end{equation}
            Figure \ref{fig-strongec} shows that only the subclasses with $K\ge 0$ obey the strong energy condition, while the the subclass with $K<0$ violates it for high values of $Q$. However, for lower values of $\beta$, one can have higher values of $Q$ than other cases without violating the strong energy condition.
	\end{enumerate}

	

	The analysis of the energy conditions in this section (figures \ref{fig-nullec}, \ref{fig-weakec}, \ref{fig-domec}, \ref{fig-strongec}), along with the causality condition (figure \ref{fig-sspeed}), regularity and finiteness of the metric functions (figures \ref{fig-e2psi}, \ref{fig-e2phi}), and monotonicity and positivity of density and pressure (figures \ref{fig-den}, \ref{fig-press}) tell us that only one subclass of the solution with $K>0$ agrees with all the basic criteria for physical acceptability \cite{delg}. Additionally, existence of an analytic, albeit complicated, expression for the equation of state makes the solution even more physically relevant.
	
	\section{Summary and Discussion} \label{sec-sumdisc}
	In this paper, we presented a comprehensive analysis of the Tolman VII solution in the presence of a charge and a cosmological constant. We started with using the Einstein-Maxwell field equations for a perfect fluid with a cosmological constant to derive a generalised TOV equation. Then, with the Tolman VII ansatz on the density profile, we solved the field equations to derive the generalised Tolman VII solution. All the features of the solution were characterised by a parameter space containing five components: the central density($\rho_0$), the self-boundedness parameter ($\beta$), the boundary radius ($r_b$), the total charge ($Q$), and the cosmological constant ($\Lambda$). Depending on the values of these parameters (specifically, the ratio, $\frac{Q^2}{\rho_0 \beta r_b^4}$), the solution can be divided into three distinct subclasses. Qualitatively these subclasses arise due to the presence of charge. This division ceases to exist with vanishing charge. Equivalently, factor that could lead to similar subclasses is the anisotropy in pressure \cite{raghoo2}.
	
	Using the metric functions, we analysed the spatial geometry of the solution and showed that for realistic values of the parameters, it remains positively curved for $0\le r\le r_b$ in all three subclasses. Then, we investigated the issue of matching the interior and exterior space-times. Due to the presence of a cosmological constant, there were two candidates to describe the (pseudo-)vacuum space-time surrounding the charged fluid sphere: the Reissner-Nordstr\"om (anti-)de Sitter solution and the charged Nariai solution. We showed that the Tolman VII solution can be matched to the Nariai solution if the matching is done at a specific boundary radius given in terms of $\beta$, $\rho_0$, $\Lambda$, and $Q$. This constituted a new class of solutions interior to the charged Nariai space-time. Interior Nariai solutions have been used to construct exact two-mass solutions \cite{bohm2,fenn}. It will be interesting to see if the new interior Nariai solution found in this paper can be used to do the same. This will be particularly significant for astrophysics since the Tolman VII solution is a reasonable model for a neutron star,and thus the two-mass solution could be used to represent a binary neutron star system.
	
	Next, we matched the interior solution with the Reissner-Nordstr\"om de Sitter space-time. The $K>0$ subclass in this case represented the generalisation of the Tolman VII solution in its usual form. Using the geodesic equations, we derived the expression for the effective gravitational potential for the fluid sphere. We also presented the general conditions for the existence of trapped null geodesics. Both $Q$ and $\Lambda$ entered these conditions. For these parameters to have any effect on geodesic trapping, we need $\Lambda \sim \rho_0 \sim \frac{Q^2}{r_b^4}$. Using such values, we presented a complete exploration of the parameter space through a series of plots. We found that the generalised solution exhibits trapped null geodesics when the tenuity ($\frac{r_b}{m(r_b)}$) is between $1.93$ and $3.06$.  This extends the bounds on tenuity from what has been previously found for the usual Tolman VII solution \cite{ishak,neary,stuch2}. This is expected due to the inclusion of $Q$ and $\Lambda$. Since the issue of geodesic trapping is of certain astrophysical relevance, it will be interesting to analyse this phenomenon with introducing rotation within our solution similar to \cite{vrba,stuch3}.
	
	Finally, we derived an analytic expressions for the pressure and the equation of state of the fluid. Both contained $\Lambda$ and $Q$ terms. The pressure was positive definite, monotonically decreasing, and vanished at the boundary for $K\ge 0$. For the class with $K<0$, the pressure became negative at certain radii, thus making this class physically irrelevant. From thereon, we focused on the subclass with $K>0$. We showed that the fluid can be described as a polytrope with an index, $\Gamma \sim 2.5$. This was consistent with previous findings \cite{raghoo}. We also showed that this subclass follows the causality criterion ($0<c_s^2<1$) and all the energy conditions. Therefore, the cosmological constant and charge (at least for reasonable values) \textit{do not} have any effect on the physical acceptability of the solution.
 
    All of the features of the solution mentioned above remain valid for a range of the self-boundedness parameter. This makes the solution all the more physically relevant since it can be used to analyse the effect of the cosmological constant on strange stars as well \cite{rej,biswas,biswas1}. It is worth noting here that the form of the charge distribution, $q(r) = \frac{Qr^3}{r_b^3}$ was chosen since it allowed for analytic solutions. Other forms could be used, however, one would have to resort to numerical methods in those cases. The solution could be further generalised by including anisotropic pressure \cite{bhar,hensh,azmat,biswas} or by deriving it in alternate/modified gravity theories \cite{stuch4,pappas,azmat,rej}. We keep these and other issues discussed above for future work.

	\section*{Acknowledgements}
	We thank Sai Madhav Modumudi and Shalini Ganguly for useful discussions. AA acknowledges partial financial support from an internal grant at St. Mary's College of Maryland.

	\appendix		

	\bibliographystyle{hunsrtnat}	
	\bibliography{tolmanvii_ext}	

\begin{thebibliography}{42}
\expandafter\ifx\csname natexlab\endcsname\relax\def\natexlab#1{#1}\fi
\expandafter\ifx\csname url\endcsname\relax
  \def\url#1{{\tt #1}}\fi

\bibitem[Misner et~al.(1973)Misner, Thorne, and Wheeler]{mtw}
Charles~W. Misner, K.S. Thorne, and J.A. Wheeler.
\newblock {\em {Gravitation}}.
\newblock W. H. Freeman, San Francisco, 1973.
\newblock ISBN 978-0-7167-0344-0, 978-0-691-17779-3.

\bibitem[Schwarzschild(1916)]{schwint}
Karl Schwarzschild.
\newblock {On the gravitational field of a sphere of incompressible fluid according to Einstein's theory}.
\newblock {\em Sitzungsber. Preuss. Akad. Wiss. Berlin (Math. Phys. )}, 1916:\penalty0 424--434, 1916, physics/9912033.

\bibitem[Tolman(1939)]{tolman1}
Richard~C. Tolman.
\newblock {Static solutions of Einstein's field equations for spheres of fluid}.
\newblock {\em Phys. Rev.}, 55:\penalty0 364--373, 1939.

\bibitem[Delgaty and Lake(1998)]{delg}
M.~S.~R. Delgaty and Kayll Lake.
\newblock {Physical acceptability of isolated, static, spherically symmetric, perfect fluid solutions of Einstein's equations}.
\newblock {\em Comput. Phys. Commun.}, 115:\penalty0 395--415, 1998, gr-qc/9809013.

\bibitem[{Raghoonundun} and {Hobill}(2015)]{raghoo}
Ambrish~M. {Raghoonundun} and David~W. {Hobill}.
\newblock {Possible physical realizations of the Tolman VII solution}.
\newblock {\em \prd}, 92\penalty0 (12):\penalty0 124005, December 2015, 1506.05813.

\bibitem[Raghoonundun et~al.(2023)Raghoonundun, Bell, and Hobill]{raghoo3}
A.~Raghoonundun, R.~Bell, and D.~Hobill.
\newblock {Exact solutions for compact stars in general relativity}.
\newblock {\em J. Phys. Conf. Ser.}, 2536\penalty0 (1):\penalty0 012003, 2023.

\bibitem[Jiang and Yagi(2019)]{jia}
Nan Jiang and Kent Yagi.
\newblock {Improved Analytic Modeling of Neutron Star Interiors}.
\newblock {\em Phys. Rev. D}, 99\penalty0 (12):\penalty0 124029, 2019, 1904.05954.

\bibitem[Posada et~al.(2021)Posada, Hlad\'\i{}k, and Stuchl\'\i{}k]{pos1}
Camilo Posada, Jan Hlad\'\i{}k, and Zden\v{e}k Stuchl\'\i{}k.
\newblock {Dynamical stability of the modified Tolman VII solution}.
\newblock {\em Phys. Rev. D}, 103\penalty0 (10):\penalty0 104067, 2021, 2103.12867.

\bibitem[Posada et~al.(2022)Posada, Hlad\'\i{}k, and Stuchl\'\i{}k]{pos2}
Camilo Posada, Jan Hlad\'\i{}k, and Zden\v{e}k Stuchl\'\i{}k.
\newblock {New interior model of neutron stars}.
\newblock {\em Phys. Rev. D}, 105\penalty0 (10):\penalty0 104020, 2022, 2201.05209.

\bibitem[Bekenstein(1971)]{bek}
Jacob~D. Bekenstein.
\newblock Hydrostatic equilibrium and gravitational collapse of relativistic charged fluid balls.
\newblock {\em Phys. Rev. D}, 4:\penalty0 2185--2190, Oct 1971.
\newblock URL \url{https://link.aps.org/doi/10.1103/PhysRevD.4.2185}.

\bibitem[Mehra(1982)]{mehra1}
A.L. Mehra.
\newblock An interior solution for a charged sphere in general relativity.
\newblock {\em Physics Letters A}, 88\penalty0 (4):\penalty0 159--161, 1982.
\newblock ISSN 0375-9601.
\newblock URL \url{https://www.sciencedirect.com/science/article/pii/0375960182905515}.

\bibitem[Florides(1983)]{flor}
P~S Florides.
\newblock The complete field of charged perfect fluid spheres and of other static spherically symmetric charged distributions.
\newblock {\em Journal of Physics A: Mathematical and General}, 16\penalty0 (7):\penalty0 1419, may 1983.
\newblock URL \url{https://dx.doi.org/10.1088/0305-4470/16/7/018}.

\bibitem[Weyl(1919)]{weyl}
H.~Weyl.
\newblock {\em Physikalische Zeitschrift}, 20:\penalty0 31--34, 1919.

\bibitem[Stuchlik(2000)]{stuch1}
Zdenek Stuchlik.
\newblock {Spherically Symmetric Static Configurations of Uniform Density in Spacetimes with a Non-Zero Cosmological Constant}.
\newblock {\em Acta Phys. Slov.}, 50\penalty0 (2):\penalty0 219--228, 2000, 0803.2530.

\bibitem[Boehmer and Fodor(2008)]{bohm}
Christian~G. Boehmer and Gyula Fodor.
\newblock {Perfect fluid spheres with cosmological constant}.
\newblock {\em Phys. Rev. D}, 77:\penalty0 064008, 2008, 0711.1450.

\bibitem[Boehmer and Mussa(2011)]{bohm2}
Christian~G. Boehmer and Atifah Mussa.
\newblock {Charged perfect fluids in the presence of a cosmological constant}.
\newblock {\em Gen. Rel. Grav.}, 43:\penalty0 3033--3046, 2011, 1010.1367.

\bibitem[Raghoonundun(2016)]{raghoo2}
Ambrish~M. Raghoonundun.
\newblock {\em {Exact Solutions for Compact Objects in General Relativity}}.
\newblock PhD thesis, Calgary U., 2016, 1604.08930.

\bibitem[Ishak et~al.(2001)Ishak, Chamandy, and Lake]{ishak}
Mustapha Ishak, Luke Chamandy, and Kayll Lake.
\newblock {Exact solutions with w modes}.
\newblock {\em Phys. Rev. D}, 64:\penalty0 024005, 2001, gr-qc/0007073.

\bibitem[Stuchl\'\i{}k et~al.(2021)Stuchl\'\i{}k, Hlad\'\i{}k, Vrba, and Posada]{stuch2}
Zden\v{e}k Stuchl\'\i{}k, Jan Hlad\'\i{}k, Jaroslav Vrba, and Camilo Posada.
\newblock {Neutrino trapping in extremely compact Tolman VII spacetimes}.
\newblock {\em Eur. Phys. J. C}, 81\penalty0 (6):\penalty0 529, 2021, 2106.05750.

\bibitem[Stephani et~al.(2003)Stephani, Kramer, MacCallum, Hoenselaers, and Herlt]{stephani}
Hans Stephani, D.~Kramer, Malcolm~A.H. MacCallum, Cornelius Hoenselaers, and Eduard Herlt.
\newblock {\em {Exact solutions of Einstein's field equations}}.
\newblock Cambridge Monographs on Mathematical Physics. Cambridge Univ. Press, Cambridge, 2003.
\newblock ISBN 978-0-521-46702-5, 978-0-511-05917-9.

\bibitem[{Goenner} and {Stachel}(1970)]{goenner}
Hubert {Goenner} and John {Stachel}.
\newblock {Einstein Tensor and 3-Parameter Groups of Isometries with 2-Dimensional Orbits}.
\newblock {\em Journal of Mathematical Physics}, 11\penalty0 (12):\penalty0 3358--3370, December 1970.

\bibitem[Oppenheimer and Volkoff(1939)]{opp}
J.~R. Oppenheimer and G.~M. Volkoff.
\newblock On massive neutron cores.
\newblock {\em Phys. Rev.}, 55:\penalty0 374--381, Feb 1939.
\newblock URL \url{https://link.aps.org/doi/10.1103/PhysRev.55.374}.

\bibitem[Mehra(1966)]{mehra}
A.~L. Mehra.
\newblock Radially symmetric distribution of matter.
\newblock {\em Journal of the Australian Mathematical Society}, 6\penalty0 (2):\penalty0 153–156, 1966.

\bibitem[Durgapal and Gehlot(1971)]{durgapal1}
M~C Durgapal and G~L Gehlot.
\newblock Spheres with varying density in general relativity.
\newblock {\em Journal of Physics A: General Physics}, 4\penalty0 (6):\penalty0 749--755, nov 1971.
\newblock URL \url{https://doi.org/10.1088/0305-4470/4/6/001}.

\bibitem[Aghanim et~al.(2020)]{planck}
N.~Aghanim et~al.
\newblock {Planck 2018 results. VI. Cosmological parameters}.
\newblock {\em Astron. Astrophys.}, 641:\penalty0 A6, 2020, 1807.06209.
\newblock [Erratum: Astron.Astrophys. 652, C4 (2021)].

\bibitem[Ray et~al.(2003)Ray, Espindola, Malheiro, Lemos, and Zanchin]{ray}
Subharthi Ray, Aquino~L. Espindola, Manuel Malheiro, Jose P.~S. Lemos, and Vilson~T. Zanchin.
\newblock {Electrically charged compact stars and formation of charged black holes}.
\newblock {\em Phys. Rev. D}, 68:\penalty0 084004, 2003, astro-ph/0307262.

\bibitem[Boehmer(2004)]{bohm3}
Christian~G. Boehmer.
\newblock {Eleven spherically symmetric constant density solutions with cosmological constant}.
\newblock {\em Gen. Rel. Grav.}, 36:\penalty0 1039--1054, 2004, gr-qc/0312027.

\bibitem[{Hidekazu}(1951)]{nari1}
Nariai {Hidekazu}.
\newblock {On a new cosmological solution of Einstein's field equations of gravitation}.
\newblock {\em Sci. Rep. Tohoku Univ. Eighth Ser.}, 35:\penalty0 46, January 1951.

\bibitem[Nariai(1999)]{nari2}
Hidekazu Nariai.
\newblock {On a New Cosmological Solution of Einstein's Field Equations of Gravitation}.
\newblock {\em Gen. Rel. Grav.}, 31\penalty0 (6):\penalty0 963--971, 1999.

\bibitem[Dev(2022)]{dev}
Krsna Dev.
\newblock {Exact solutions for charged spheres and their stability. I. Perfect Fluids}.
\newblock 2 2022, 2202.07819.

\bibitem[Neary et~al.(2001)Neary, Ishak, and Lake]{neary}
Nicholas Neary, Mustapha Ishak, and Kayll Lake.
\newblock {The Tolman VII solution, trapped null orbits and W modes}.
\newblock {\em Phys. Rev. D}, 64:\penalty0 084001, 2001, gr-qc/0104002.

\bibitem[Fennen and Giulini(2015)]{fenn}
Michael Fennen and Domenico Giulini.
\newblock {Static and spherically symmetric two-mass solutions of Einstein-Maxwell equations with cosmological constant based on Nariai spacetime}.
\newblock {\em Class. Quant. Grav.}, 32\penalty0 (4):\penalty0 045008, 2015, 1408.2713.

\bibitem[Vrba et~al.(2020)Vrba, Urbanec, Stuchl\'\i{}k, and Miller]{vrba}
Jaroslav Vrba, Martin Urbanec, Zden\v{e}k Stuchl\'\i{}k, and John~C. Miller.
\newblock {Trapping of null geodesics in slowly rotating spacetimes}.
\newblock {\em Eur. Phys. J. C}, 80\penalty0 (11):\penalty0 1065, 2020, 2011.13616.

\bibitem[Stuchl\'\i{}k and Vrba(2021)]{stuch3}
Zden\v{e}k Stuchl\'\i{}k and Jaroslav Vrba.
\newblock {Trapping of null geodesics in slowly rotating extremely compact Tolman VII spacetimes}.
\newblock {\em Eur. Phys. J. Plus}, 136\penalty0 (9):\penalty0 977, 2021, 2108.09466.

\bibitem[Rej et~al.(2023)Rej, Errehymy, and Daoud]{rej}
Pramit Rej, Abdelghani Errehymy, and Mohammed Daoud.
\newblock {Charged strange star model in Tolman\textendash{}Kuchowicz spacetime in the background of 5D Einstein\textendash{}Maxwell\textendash{}Gauss\textendash{}Bonnet gravity}.
\newblock {\em Eur. Phys. J. C}, 83\penalty0 (5):\penalty0 392, 2023, 2305.06748.

\bibitem[Biswas et~al.(2019)Biswas, Shee, Ray, Rahaman, and Guha]{biswas}
Suparna Biswas, Dibyendu Shee, Saibal Ray, F.~Rahaman, and B.~K. Guha.
\newblock {Relativistic Strange Stars in Tolman-Kuchowicz Spacetime}.
\newblock {\em Annals Phys.}, 409:\penalty0 167905, 2019, 1910.00427.

\bibitem[Biswas et~al.(2020)Biswas, Shee, Guha, and Ray]{biswas1}
Suparna Biswas, Dibyendu Shee, B.~K. Guha, and Saibal Ray.
\newblock {Anisotropic strange star with Tolman\textendash{}Kuchowicz metric under $f(R,T)$ gravity}.
\newblock {\em Eur. Phys. J. C}, 80\penalty0 (2):\penalty0 175, 2020, 2006.01619.

\bibitem[{Bhar} et~al.(2015){Bhar}, {Murad}, and {Pant}]{bhar}
Piyali {Bhar}, Mohammad~Hassan {Murad}, and Neeraj {Pant}.
\newblock {Relativistic anisotropic stellar models with Tolman VII spacetime}.
\newblock {\em Astrophysics and Space Science}, 359:\penalty0 13, September 2015.

\bibitem[Hensh and Stuchl\'\i{}k(2019)]{hensh}
Sudipta Hensh and Zden\v{e}k Stuchl\'\i{}k.
\newblock {Anisotropic Tolman VII solution by gravitational decoupling}.
\newblock {\em Eur. Phys. J. C}, 79\penalty0 (10):\penalty0 834, 2019, 1906.08368.

\bibitem[Azmat and Zubair(2021)]{azmat}
Hina Azmat and M.~Zubair.
\newblock {An anisotropic version of Tolman VII solution in $f(R, T)$ gravity via gravitational decoupling MGD approach}.
\newblock {\em Eur. Phys. J. Plus}, 136\penalty0 (1):\penalty0 112, 2021, 2106.08384.

\bibitem[Stuchlik et~al.(2011)Stuchlik, Hladik, and Urbanec]{stuch4}
Zdenek Stuchlik, Jan Hladik, and Martin Urbanec.
\newblock {Neutrino trapping in braneworld extremely compact stars}.
\newblock {\em Gen. Rel. Grav.}, 43:\penalty0 3163--3190, 2011, 1108.5767.

\bibitem[Pappas et~al.(2022)Pappas, Posada, and Stuchl\'\i{}k]{pappas}
Thomas~D. Pappas, Camilo Posada, and Zden\v{e}k Stuchl\'\i{}k.
\newblock {Extended Tolman III and VII solutions in f(R,T) gravity: Models for neutron stars and supermassive stars}.
\newblock {\em Phys. Rev. D}, 106\penalty0 (12):\penalty0 124014, 2022, 2210.15597.

\end{thebibliography}
	\addcontentsline{toc}{section}{References}	
	
\end{document}